\documentclass[journal]{IEEEtran}
\usepackage{cite}
\usepackage{amsmath,amssymb,amsfonts}
\usepackage{bbm}
\usepackage{soul}
\usepackage{graphicx}
\usepackage{textcomp}
\usepackage{xcolor}
\usepackage{tikz}
\newcommand\checkmarks[1][]{%
  \tikz[scale=0.4,#1]{\fill(0,.35) -- (.25,0) -- (1,.7) -- (.25,.15) -- cycle;}%
}
\newcommand\crossmark[1][]{%
  \tikz[scale=0.4,#1]{
    \fill(0,0)--(0.1,0) .. controls (0.5,0.4) .. (1,0.7)--(0.9,0.7) ..  controls (0.5,0.5) ..(0,0.1) --cycle;
    \fill(1,0.1)--(0.9,0.1) .. controls (0.5,0.3) .. (0,0.7)--(0.1,0.7) .. controls (0.5,0.4) ..(1,0.2) --cycle;
  }%
}
\usepackage{cancel}
\usepackage{kotex}
\usepackage{circledsteps}
\usepackage{balance}
\usepackage[normalem]{ulem}
\usepackage{tikz}
\usepackage{subfigure}
\usepackage{booktabs}
\usepackage{color,bbm}
\usepackage{tabularx}
\usepackage[linesnumbered,ruled]{algorithm2e}
\newcommand{\argmaxD}{\arg\!\max} 
\newcommand*\circled[1]{\Circled[inner color=white, fill color= color0, outer color=color0]{\footnotesize{#1}}} 
 
\newcommand{\BfPara}[1]{{\noindent\bf#1.}\xspace}

\newcommand{\revision}[1]{{\color{black}{#1}}}
\newcommand{\rev}[1]{{\color{black}{#1}}}

\definecolor{color1}{RGB}{228,26,28}
\definecolor{color2}{RGB}{55,126,184}
\definecolor{color3}{RGB}{77,175,74}
\definecolor{color4}{RGB}{152,78,163}
\definecolor{color5}{RGB}{255,127,0}
\definecolor{color6}{RGB}{241,241,241}
\definecolor{color7}{RGB}{156,156,156}
\definecolor{color8}{RGB}{96,96,96}
\definecolor{color0}{RGB}{162, 20, 47}
\usetikzlibrary{patterns,arrows}
\def\BibTeX{{\rm B\kern-.05em{\sc i\kern-.025em b}\kern-.08em
    T\kern-.1667em\lower.7ex\hbox{E}\kern-.125emX}}
    
\begin{document}
\providecommand{\keywords}[1]{\textbf{\textit{Index terms---}} #1}
\title{Quantum Multi-Agent Actor-Critic Networks for Cooperative Mobile Access in Multi-UAV Systems}
\author{
    Chanyoung Park, 
    Won Joon Yun, 
    Jae Pyoung Kim, 
    Tiago Koketsu Rodrigues, 
    Soohyun Park, 
    \\
    Soyi Jung,~\IEEEmembership{Member,~IEEE}, 
    and
    Joongheon Kim,~\IEEEmembership{Senior Member,~IEEE}
    \thanks{Parts of this research were appeared at \textit{IEEE International Conference on Distributed Computing Systems (ICDCS)}, Bologna, Italy, July 2022~\cite{yun2022quantum}.}
    \thanks{This work was supported by the JSPS/NRF/NSFC A3 Foresight Program. \textit{(Corresponding authors: Soohyun Park, Soyi Jung)}} 
    \thanks{C. Park, W.J. Yun, J.P. Kim, S. Park, and J. Kim are with School of Electrical Engineering, Korea University, Seoul 02841, Korea (e-mails: \{cosdeneb,ywjoon95,paulkim436,soohyun828,joongheon\}@korea.ac.kr).}
    \thanks{T.K. Rodrigues is with Graduate School of Information Sciences, Tohoku University, Sendai, Japan (e-mail: tiago.gama.rodrigues@it.is.tohoku.ac.jp).}
    \thanks{S. Jung is with Department of Electrical of Computer Engineering, Ajou University, Suwon 16499, Korea (e-mail: sjung@ajou.ac.kr).}
}
\maketitle

\begin{abstract}
This paper proposes a novel algorithm, named quantum multi-agent actor-critic networks (QMACN) for autonomously constructing a robust mobile access system employing multiple unmanned aerial vehicles (UAVs). In the context of facilitating collaboration among multiple unmanned aerial vehicles (UAVs), the application of multi-agent reinforcement learning (MARL) techniques is regarded as a promising approach. These methods enable UAVs to learn collectively, optimizing their actions within a shared environment, ultimately leading to more efficient cooperative behavior. Furthermore, the principles of a quantum computing (QC) are employed in our study to enhance the training process and inference capabilities of the UAVs involved. By leveraging the unique computational advantages of quantum computing, our approach aims to boost the overall effectiveness of the UAV system. However, employing a QC introduces scalability challenges due to the near intermediate-scale quantum (NISQ) limitation associated with qubit usage. The proposed algorithm addresses this issue by implementing a quantum centralized critic, effectively mitigating the constraints imposed by NISQ limitations. Additionally, the advantages of the QMACN with performance improvements in terms of training speed and wireless service quality are verified via various data-intensive evaluations. Furthermore, this paper validates that a noise injection scheme can be used for handling environmental uncertainties in order to realize robust mobile access.
\end{abstract}


\section{Introduction}\label{sec:1}
\subsection{Background and Motivation}
High-mobility, agility, and accessibility are the representative advantages of unmanned aerial vehicles (UAVs) that help serve a wide range of services beyond 5G or 6G~\cite{9044378,9428629,9459481,9023459,saad2019vision}. UAV deployments continue to grow, and the number of UAV fleets will reach 1.6 million by 2024 according to a Federal Aviation Administration (FAA) prediction~\cite{mozaffari2021toward}. Examples of UAV applications include real-time surveillance~\cite{yun2022cooperative}, package delivery~\cite{9098900}, smart factory~\cite{10012051,lee2018online}, and disaster monitoring/management~\cite{9583845}. Notably, UAVs provide wireless communication service without a fixed terrestrial infrastructure to create mobile access networks, providing broader wireless coverage and real-time service~\cite{tvt202106jung,8660516}.
UAVs can rapidly establish wireless connections with ground users in emergencies such as military issues or disasters and in extreme areas where installing terrestrial base stations is not easy for technical or economic reasons~\cite{zhang2021joint, bor2016new,zeng2016wireless}.
Developing a mobile access system with the aid of UAVs is a challenging task due to the multitude of unanticipated variables and uncertainties in a practical setting. These variables may include obstacles, strong winds, collisions, limited energy reserves, and potential malfunctions.
Hence, designing a machine learning algorithm capable of autonomously and adaptively determining optimal UAV trajectories serves as an alternative approach to anticipating and responding to the abovementioned environmental uncertainties.

Among several machine learning (ML) and deep learning (DL) algorithms, reinforcement learning (RL) has exhibited the most remarkable performance in tasks requiring sequential decision-making processes~\cite{pieee202105park}.
RL enables effective resource management service in a dynamic networking system, including UAV-aided mobile access~\cite{lahmeri2021artificial}. In addition, RL is also scalable such that multi-agent reinforcement learning (MARL) can be used to control multiple UAVs. However, within the MARL architecture, various interactions occur between UAVs which will help train their policies contemporaneously and affect each other's policy training performance. In addition, UAVs observe partial information about the dynamic environment due to physical limitations. These factors make MARL environment non-stationary. 

Quantum MARL (QMARL) using quantum entanglement can handle the above non-stationary problem~\cite{oh2020tutorial}. Building quantum neural networks (QNNs) using quantum computing (QC) enables efficient consumption of computing resources with fewer model parameters than conventional neural networks, resulting in more reliable and faster learning~\cite{chen2020variational, yun2022quantum,9555241}. 
Existing studies have performed centralized training and decentralized execution (CTDE)-based computation for the cooperation of multiple agents using a structure called \textit{CommNet}~\cite{sukhbaatar2016learning}. \textit{CommNet} is a structure that receives state information of all agents as input and learns the actions of all agents in one single neural network. If $m$ agents represent the state as an $n$-dimensional vector, the input of the CommNet increases to $m\times n$. Accordingly, the input size scales up linearly as the number of agents increases. However, this linear scale-up is very burdensome in a QC environment due to the near intermediate-scale quantum (NISQ) limitation in using qubits, which causes scalability issues~\cite{shor1995scheme}.

Quantum computers manufactured by companies such as \textit{IBM} and \textit{Google} have massive sizes. Thus, the idea of quantum UAVs may seem implausible and unrealistic. However, they are rapidly becoming feasible because of the advancements in \textit{miniature quantum computers}. Until now, there are several successful models of small quantum computers.
For instance, \textit{Gen1} miniature quantum computer, developed by \textit{Quantum Brilliance}, utilizes 5 qubits and possesses dimensions suitable for a rack-mountable unit~\cite{horsley_2022, liu2022nitrogen}. Its compact size is notably smaller than conventional quantum computers, making it suitable to be placed on a desk or tabletop.
Next, \textit{SpinQ} has also succeeded in creating an even smaller quantum computer known as \textit{Gemini mini}~\cite{xiang_2022}. Although it can only handle 2 qubits, it is even smaller than the work of \textit{Quantum Brilliance}, making it much more compatible with UAVs. 
Lastly, a rack-sized 20 qubits quantum computer was produced by \textit{Alpine Quantum Technologies}~\cite{blatt_monz_zoller_2022}.
Even though the aforementioned quantum computer cannot be mounted on a UAV, its importance lies in demonstrating that the size of quantum computers can be greatly decreased while still preserving the number of usable qubits.
As seen from the examples above, the number of available qubits is directly related to the size of quantum computers and miniature quantum computers inevitably have limited performance. Yet, this research is still an ongoing process and the scale of qubit reduction is decreasing as the technology continues to advance. 
Hence, the implementation of quantum UAVs employing high-functioning, small-scale quantum computers is definitely a feasible concept in the near future, bolstering the practicality of our work.
\rev{Based on the advances in QC and QMARL, the related deep learning and AI technologies can also be utilized for the performance improvements of multi-UAV autonomous networks in terms of mobility control, network scheduling, and so forth. However, UAV operation is highly susceptible to variations caused by the environment (e.g., weather affecting UAV movement), which can lead to a significant amount of noise in the corresponding data. The design of QC and QMARL solutions for UAV systems thus should take this noise into account and be customized to specifically handle these UAV characteristics.}

\subsection{Algorithm Design Rationale}
\begin{figure}
    \centering
    \includegraphics[width=\linewidth]{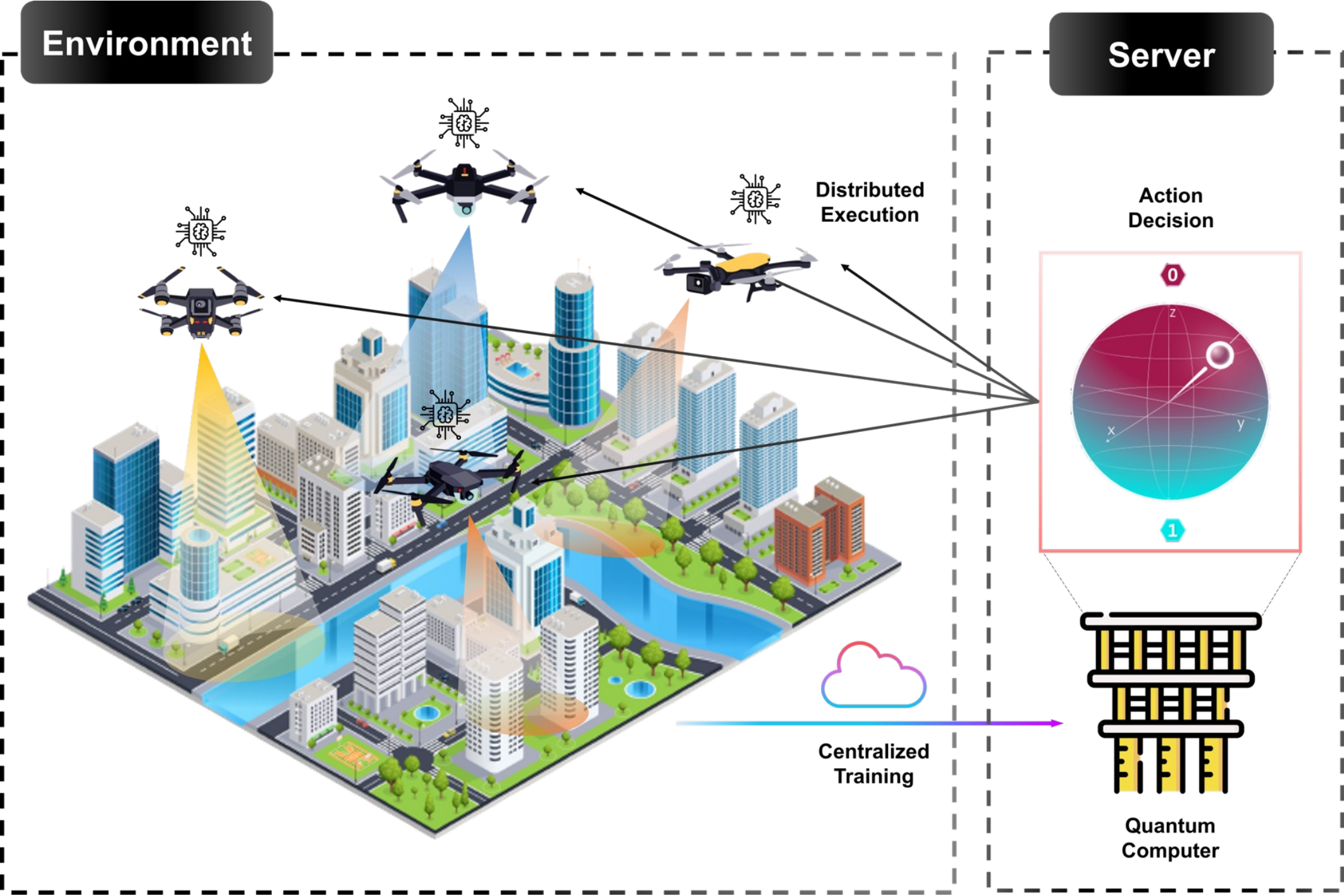}
    \caption{Proposed mobile access architecture for quantum multi-UAV systems.}
    \label{fig:sys_architecture}
\end{figure}

We address the scalability issue related to quantum error arising from increasing the number of employed qubits, as previously mentioned, in the quantum domain by using a structure in which a \textit{centralized critic} network trains multiple \textit{actor} networks iteratively~\cite{yun2022quantum}.
This paper proposes a CTDE-based training algorithm, referred to as a quantum multi-agent actor-critic networks (QMACN) algorithm utilizing the benefit of a QC (\textit{i.e.}, quantum entanglement, efficient utilization of computing resource with fewer parameters).
As observed in the illustration presented in Fig.~\ref{fig:sys_architecture}, the proposed algorithm has been utilized to develop a self-governing mobile connectivity infrastructure employing UAVs within the scope of the study. Within the suggested paradigm, a \textit{centralized critic} situated in the main server appraises each UAV's decision-making capabilities throughout the training procedure. Subsequent to the completion of the centralized training phase, an array of UAVs collaboratively deliver wireless communication services to terrestrial users by employing their learned parameterized strategies. With the proposed CTDE-centric QMARL algorithm, inter-UAV communications are not needed thanks to the information-sharing by employing a \textit{centralized critic}.

Finally, this paper designs realistic environments by taking into account the environmental information noise affecting UAVs, which usually hinders them from behaving as intended.
Considering environmental factors, this paper corroborates UAVs react more adaptively to environmental noises through performance evaluations.

\subsection{Contributions} 
The contributions of our work are as follows.
\begin{itemize}
    \item This paper presents a \rev{novel algorithm, referred to as} {QMACN} enabling multiple UAVs to provide autonomous mmWave communication services cooperatively by training UAVs to find optimal trajectories in uncertain and dynamic environments.
    \item A novel CTDE framework is proposed to \revision{implement} QMARL \revision{while mitigating the constraints} of quantum computing, \textit{i.e.}, quantum error\revision{s} \revision{arising from} increasing the number of qubits \revision{during} the NISQ era.
    \item This paper designs a realistic environment by noise injection to construct robust mobile access in multi-UAV systems, where noises are considered based on real environments. A representative characteristic of RL is to train agents to respond adaptively to dynamic elements.
    \item This paper corroborates the advantages of QMARL and realistic noise reflection on training policies by carrying out various data-intensive evaluations in policy training and its inference process.
    \item \revision{While the prior works in~\cite{yun2022quantum, 10012051} also suggest the QMACN architecture, their applications primarily target one-hop scheduling and smart factories, respectively. In contrast, the current work emphasizes the development of an autonomous mobile access system, taking into account specific UAV specifications and a realistic setting. Moreover, we demonstrate quantum supremacy more comprehensively than the aforementioned works, utilizing experimental results obtained through noise injection or mathematical computational complexity analysis.}
\end{itemize}

\subsection{Organization} 
The rest of this paper is organized as follows\revision{:}
Sec.~\ref{sec:2} introduces several articles of related work and technologies.
Sec.~\ref{sec:3} reviews preliminary knowledge. The details of the baseline models are described in Sec.~\ref{sec:4}. Sec.~\ref{sec:5} presents a novel \revision{QMARL} algorithm for multi-UAV mobile access. Sec.~\ref{sec:6} evaluates the performance via data-intensive simulations. Lastly, Sec.~\ref{sec:7} concludes this paper.

\section{Related Work}\label{sec:2}

\subsection{Mobile Cellular Access via UAV Networks}
\revision{A considerable body of research has been published on the optimization of UAV deployment, trajectory, and path planning in cellular services, with the aim of improving wireless communication systems. The research in~\cite{mozaffari2016optimal} aims to minimize transmission power in UAV deployment, taking into account various uncertainties such as fluctuations in user demands. System sum capacity is enhanced through a combination of K-means and Q-learning in~\cite{deployment4}, and both uplink and downlink interferences are considered where ground and aerial base stations coexist~\cite{deployment5}. In~\cite{deployment1}, a 3D multi-UAV deployment strategy is proposed, ensuring quality of service (QoS) requirements for diverse user distributions while taking co-channel interference into account. The objective of~\cite{kalantari2016number} is to maximize coverage with the fewest UAV base stations possible. In addition, the work in~\cite{cheng2018uav} investigates optimal path planning for multiple UAVs to deliver wireless services to cell edge users using a convex relaxation technique. To accommodate moving users, the positioning of airship-based flying base stations (FlyBSs) is analyzed in relation to the balance between total capacity and energy consumption~\cite{deployment2}. Lastly, a circle packing-based algorithm is examined for optimal quasi-stationary deployment systems, aiming to maximize overall throughput~\cite{deployment3}.
Although prior studies demonstrate satisfactory performance concerning their objectives, their solution approaches are focused on centralized optimization problems. Thus, the aforementioned methods fall short in offering real-time solutions for highly dynamic and massive UAV-aided networks. In order to tackle these challenges, our paper introduces an approach founded on the CTDE-based MARL framework, which is effective in addressing the highlighted challenges within a distributed manner.}

\revision{Subsequently, there are studies that employ a QC to propose quantum UAVs for the establishment of communication networks, thereby demonstrating the feasibility of quantum UAVs.}
In the first work, quantum UAVs generate and transmit entangled photon pairs to transfer information which is feasible because it does not require a highly advanced quantum computer. Initially, a simple network composed of only a pair of quantum UAVs is formed to verify the concept~\cite{liu2020drone}.
Afterwards, this work was further expanded by increasing the number of quantum UAVs and the size of the quantum communication network~\cite{liu2021optical}. Despite the increased size of the network and unstable entangled photons, this work was able to successfully transmit data accurately from one point to another. 
Finally, a research on utilizing quantum UAVs in a metropolitan environment was done as well~\cite{chiti2022metropolitan}.
A swarm of quantum UAVs exist in the scenario of this work and they are used to build a quantum communication network by acting as signal repeater nodes. Thus, when a ground-based quantum server transmits data via photons, the swarm of quantum UAVs amplify the data signal such that it is transmitted over a long distance. As a result, loss of data is minimized and an efficient quantum communication network is implemented. 
From these works, we verify that quantum UAVs are feasible and can be practically implemented. Furthermore, tasks much more complex will be achieved in the near future as miniature quantum computers continue to advance.

\subsection{Applications of MARL with UAVs}
UAVs have the potential to augment the capabilities of MARL systems in a wide range of applications, by offering aerial perspectives, coverage, and mobility.
A fleet of UAVs trained through MARL can undertake surveillance and monitoring tasks, such as target tracking or area monitoring for security purposes~\cite{tii202210yun,xia2021multi,baldazo2019decentralized}.
UAVs equipped with sensors and cameras can collect data on environmental factors like air quality, temperature, and humidity~\cite{cai2021cooperative,oubbati2021multi,tang2022incentivizing}. A team of UAVs can be trained using MARL to coordinate and cover extensive areas for data collection.
Moreover, a group of UAVs can be trained through MARL to collaborate on search and rescue missions in disaster-stricken areas, identifying survivors, assessing damages, and offering assistance~\cite{sadhu2020aerial,gomez2022leader}.
Finally, UAVs can establish temporary communication networks in remote or disaster-impacted regions. A group of UAVs trained with MARL can function as relays, passing messages among themselves until they reach their intended destination~\cite{tan2022cooperative,albinsaid2021multi,wang2022learning}.
In our research work, the aim is to harness the quantum supremacy of a QC, which is defined by a lower count of parameters and diminished computational complexity compared to traditional neural network-driven MARL learning, within the realm of UAV applications.

\rev{
\subsection{Previous Work}
Our previous QMARL applications to network and mobile systems are discussed in \cite{yun2022quantum,10012051}.
In~\cite{yun2022quantum}, a parallel single-hop network scheduling algorithm is designed and demonstrated. The use of QMARL for this single-hop network scheduling can achieve significant performance improvements. Obviously, this network scheduling formulation is fundamentally different from our considering multi-UAV mobility control. 
In~\cite{10012051}, QMARL-based mobility control for autonomous mobile vehicles is discussed. Even if the proposed algorithm in~\cite{10012051} also considers mobility control, it is not same with the proposed algorithm in this paper because this paper includes (i) practical noise factors considerations in actions and states which are challenging in MARL computation, (ii) computational complexity analysis, (iii) UAV energy model consideration which is not conducted in~\cite{10012051}, and lastly, (iv) aerial environments and tasks construction which are totally different from smart factory environments and tasks~\cite{10012051}. 
}

\section{Quantum Machine Learning}\label{sec:3}

\begin{figure}
    \centering
    \subfigure[Classical \revision{Neural Network}.]
    {
    \includegraphics[width=0.95\linewidth]{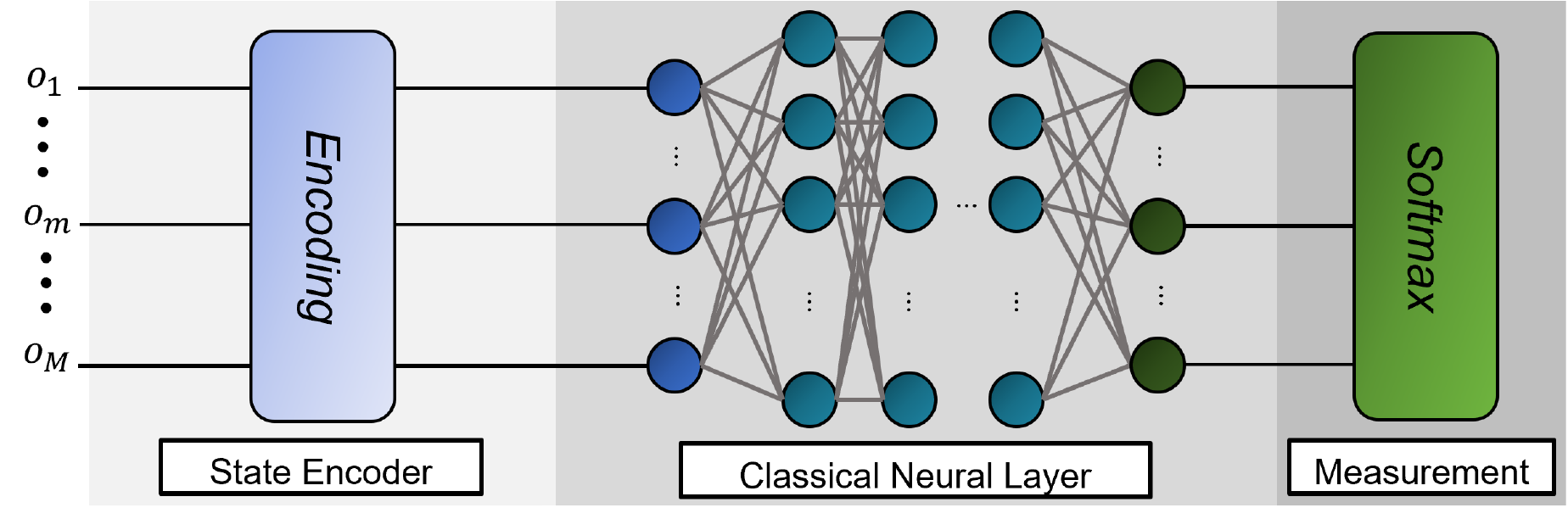}
    \label{fig:MARL_Layer}
    }
    \subfigure[Quantum \revision{Neural Network}.]
    {
    \includegraphics[width=0.95\linewidth]{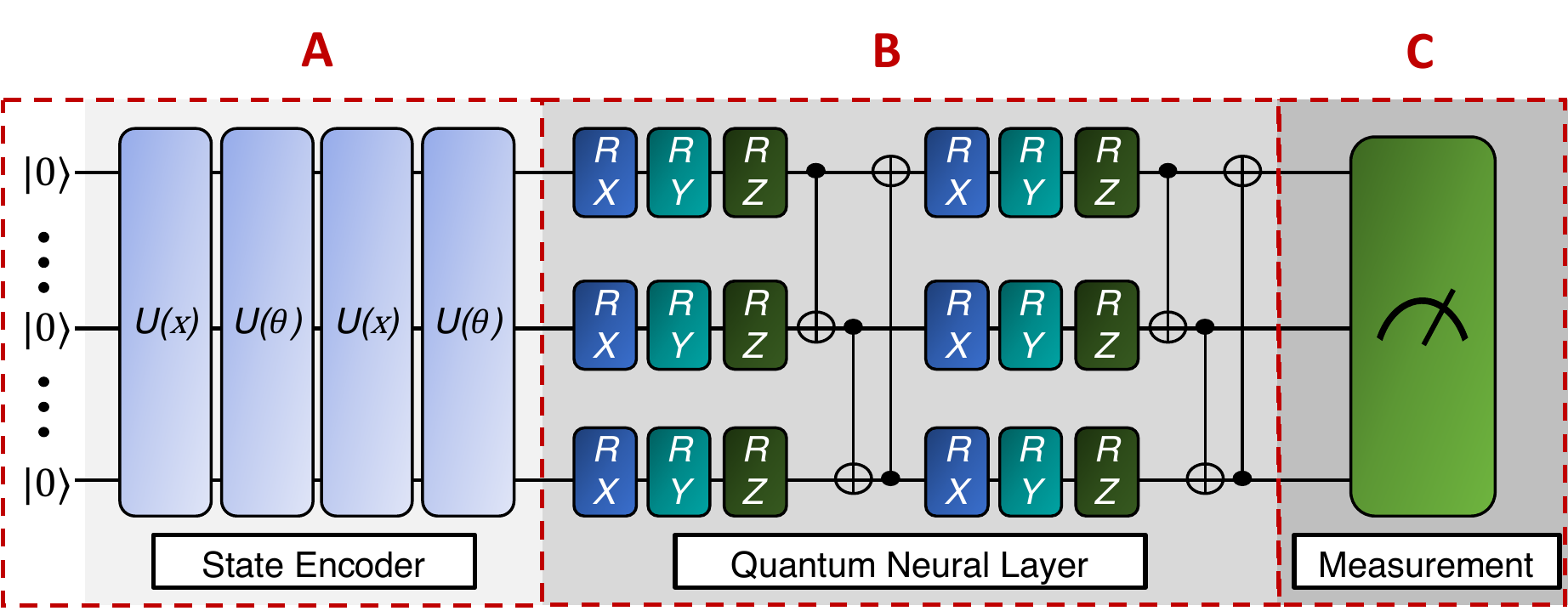}
    \label{fig:Quantum_Layer}
    }
    \caption{Architecture of \revision{the neural network based on} classical computing and quantum computing \revision{for reinforcement learning}.}
\end{figure}

\subsection{Qubits}
Qubits are used in \revision{a} QC as the basic unit of information. They can take any value between 0 and 1 because they are expressed as a combination of two bases: $|0\rangle$ and $|1\rangle$. A qubit that is composed of both bases is said to be in a state of \textit{superposition} and can also be used to express quantum states~\cite{bouwmeester2000physics}. In addition, \textit{entanglement} between qubits is also possible, which significantly increases the correlation between two individual qubits. This nature of qubits allows \revision{the} QC to contain and control more information compared to classical computing. Assuming a $q$ qubits system, a quantum state existing in the Hilbert state of the system is expressed as $|\psi\rangle=\alpha_{1}|0\cdots 0\rangle + \cdots + \alpha_{2^q}|1\cdots 1\rangle$,
where $\alpha$ stands for the probability amplitudes of each base, \textit{i.e.}, $\sum^{2^q}_{i=1}\alpha_{i}^2=1$ and $\psi$ represents the quantum state.
As $\alpha$ is a complex number, a quantum state is expressed as a point on the Bloch sphere. 

\subsection{Quantum Neural Network}
In order to design and compute QNN using qubits, the qubits should be controllable for training the neural network. The control is achieved via the utilization of basic quantum gates in order to control the positions of qubits over Bloch sphere. Representative examples of the basic quantum gates are rotation gates, which are expressed as $R_x$, $R_y$, and $R_z$, which are for the rotation over $x$-, $y$-, and $z$-axes. For more details, the gate functions are performed as unitary operations on a single qubit, causing it to rotate by a specific value in the given directions of $x$-, $y$-, and $z$-axes. \revision{T}hese gates not only control qubits but also encode classical bit-scale data. While basic quantum rotation gates are single qubit gates that can only be applied to a single qubit simultaneously, there are also multiple qubit gates acting on two or more qubits simultaneously. For example, a \textit{CNOT} gate causes entanglement among several qubits by performing an \textit{XOR} operation on two qubits~\cite{williams1998explorations}. 

Based on \revision{the above} theories and concepts, QNN models are built by assembling \revision{various types of} gates. Conventional QNN models consist of following three components, \textit{i)} state encoding circuit (refer to \textsf{A} in Fig.~\ref{fig:Quantum_Layer}), \textit{ii)} parameterized quantum circuit (PQC) (refer to \textsf{B} in Fig.~\ref{fig:Quantum_Layer}), and \textit{iii)} quantum measurement (refer to \textsf{C} in Fig.~\ref{fig:Quantum_Layer}) layers.
\revision{Further description regarding the three consecutive components can be found in the subsequent sections.}

\subsubsection{State Encoding Circuit}
First, the encoding layer's function is to encode classical data into quantum states because quantum circuits cannot take classical bits as input. Therefore, the state encoder converts bits into qubits by passing $q$ number of $|0\rangle$ into an array of rotation gates using classical data used as parameters denoted as $\theta_{enc}$. Additionally, the input data $X$ is split into $[x_{1} \cdots x_{N}]$ such that they can be individually used as parameters, where $N$ is the number of split data for the input $X$. Then, the output quantum state of \revision{the encoding} layer will contain the information of classical data. 


\subsubsection{Parameterized Quantum Circuit}
Secondly, there is PQC which carries out the desired computation, and it is equivalent to a classical neural network (NN), especially accumulated hidden layer multiplication \revision{as depicted in Fig.~\ref{fig:MARL_Layer}}. In \revision{the} layer of \revision{PQC}, the input quantum state is rotated by a specific angle using quantum gates such that the output will give the required value like the action and state values. \revision{In our} paper, the qubits are computed using the \textit{Controlled-Universal} (CU) gate which has flexible control over the direction of rotation, entanglement, and disentanglement.  
The structure of the QNN model is \revision{presented} in Fig.~\ref{fig:Quantum_Layer}, and it can be seen that the encoding layer followed by the CU3 layer is repeated several times. \revision{The} particular structure is due to the data re-uploading technique~\cite{P_rez_Salinas_2020}, simultaneously encoding and rotating the qubits. As a result, the computation efficiency of each qubit is maximized, \textit{i.e.}, the number of qubits is decreased which is required to produce the values needed for MARL.

\subsubsection{Quantum Measurement}
Lastly, the quantum state produced from PQC becomes the input of the measurement layer. In this stage, the input is measured such that the quantum data are decoded back into classical data for optimization. \revision{The} measurement operation is equivalent to the multiplication of a projection matrix with respect to $z$-axis. While the $z$-axis is most commonly used for measurement, it can be any other properly defined directions. After conducting the measurement of the quantum state, the quantum state collapses, and it becomes an \textit{observable}. 
After the decoding procedure, the \textit{observable} is used to minimize the loss function. Then, it should be differentiated for backpropagation. However, quantum data cannot be differentiated because applying \revision{the} chain rule will completely collapse the \revision{state of} qubits. Thus, the technique for obtaining the loss gradient via the symmetric difference quotient of the loss function of the observable is used for QNN training.

\subsection{Quantum Reinforcement Learning}
The quantum circuit's final \textit{observable} value represents an agent's action probability in MARL computation. It coincides with the \textit{softmax} function, as illustrated in Fig.~\ref{fig:MARL_Layer}. Using the information in the replay buffer will compute the action distribution of agents in the environment. Then, as \revision{the} agent \revision{makes decision-making} based on this action, a new state, observation, and reward data will be produced for another repeated iteration of learning.
Previous studies have already verified that quantum circuits show better performance than DNN-based RL methods composed of the same parameters~\cite{lockwood2020reinforcement, chen2020variational}. 
\revision{
The study in~\cite{yun2022quantum} represents one of the initial attempts to develop a purely quantum version of classical MARL algorithms. Through conducting several simulations in a single-hop data offloading environment, the research confirmed that QMARL surpassed its classical equivalent by capitalizing on the advantages of a QC. Subsequently, the research in~\cite{10012051} presented a novel contribution by implementing the aforementioned algorithm in a smart factory context. While no substantial enhancements were made to the algorithm itself, the work validated the practical effectiveness of the quantum algorithm through meticulous simulations.
In addition, the potential of leveraging a distinct quantum attribute referred to as pole memory is explored for conducting meta-learning~\cite{yun2022quantum2}. It reveals that quantum reinforcement learning (QRL) executes multiple tasks through few-shot learning, rendering QRL highly efficient.}
The algorithm in \revision{our} paper takes advantage of \revision{a} QC for \revision{resilient} mobile access\revision{,} and it investigates the performance \revision{assessment} of QMARL \revision{in} \revision{comparison} to traditional MARL\revision{, demonstrating} quantum supremacy \revision{for the problems considered in our study}.
\revision{Although the previous studies in~\cite{yun2022quantum, 10012051} also utilize the QMARL framework, their applications mainly focus on one-hop scheduling and smart factories, respectively. Furthermore, our work provides a more thorough demonstration of quantum supremacy compared to the mentioned studies, by employing experimental results derived from noise injection in Sec.~\ref{sec:6} and mathematical computational complexity analysis in Sec.~\ref{sec:5-c}.}

\begin{figure*}[t]
    \centering
    \includegraphics[width=\linewidth]{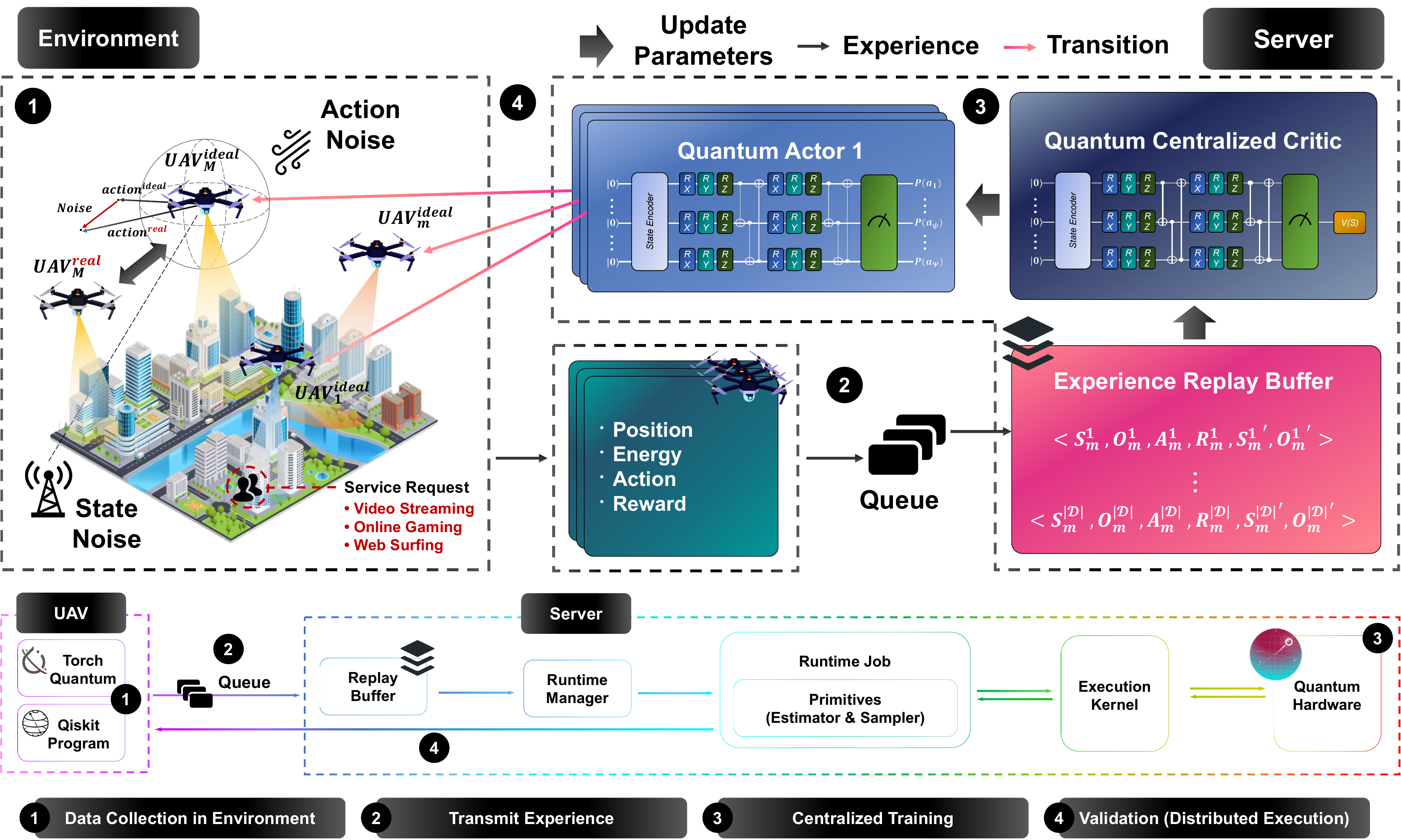}
    \caption{\revision{Proposed CTDE-based training pipeline for constructing autonomous quantum-based multi-UAV networks.}}
    \label{fig:sys_pipeline}
\end{figure*}

\section{Model}\label{sec:4}
\subsection{Mobile Access Model using Multi-UAV Networks}
We propose \revision{an algorithm called} QMACN aiming to construct reliable and robust autonomous multi-UAV networks to use QMARL in a dynamic environment, as illustrated in Fig.~\ref{fig:sys_pipeline}.
\revision{As detailed in Sec.~\ref{sec:1}, the proposed algorithm is implemented in the context of a UAV-supported autonomous mobile access system, addressing the scalability challenges from the excessive employment of qubits in the NISQ era.
Within the schematic presented in Fig.~\ref{fig:sys_pipeline}, a multitude of UAVs, which are \textit{actors}, garner environmental experiences encompassing their locations, residual energy levels, executed actions, and acquired rewards while providing wireless communication services to terrestrial users.
Then, these UAVs convey their experiences to a central server. In the context of the paper, the server's queue state is assumed to be ideal, resulting in the absence of any delays.}
It also takes into account the situation where if communication for sending a UAV's experience fails, it will retransmit until it succeeds like the standard transmission control protocol (TCP).
\revision{Within the central server, a \textit{centralized critic} facilitates the learning of parameters for multiple \textit{actors} by evaluating the decision-making competencies of each UAV under specific states throughout the comprehensive training procedure.
Following the conclusion of the centralized training phase, a collection of UAVs cooperatively provide wireless communication services to ground users based on their learned policies in a decentralized manner.}
Considering the CTDE environment, UAVs may encounter multifaceted uncertainties \revision{manifested} as \revision{environmental} noise when sending their experiences to the central server \revision{for training purposes}. Not only do the interactions between UAVs affect the training policies of other agents, but various noise factors do affect them as well. It is the reason for using RL, and two types of noises are fed into our MARL system to consider a realistic environment in this paper; \textit{i)} state noise and \textit{ii)} action noise. 


\subsection{UAV Model}
\revision{As UAVs are equipped with onboard batteries, their operations are subject to a restricted battery capacity~\cite{9447255}.
In instances where UAVs malfunction due to the battery being completely discharged, they might fall to the ground.
This scenario presents the possibility of causing physical harm to other UAVs or ground users, ultimately undermining the trustworthiness of the services delivered. Hence, monitoring and managing the residual energy of UAVs is crucial in ensuring the services' reliability~\cite{mei2019joint}.}
In general, the main sources of energy consumption for UAVs in mobile access lie in aviation and wireless communications. However, this paper only considers energy consumption by aviation because communication-related energy (a few \revision{W}atts) is much lower than aviation-related energy (a few hundred \revision{W}atts)~\cite{tvt202106jung}.
The energy expenditure model of \revision{the} $m$-th UAV when hovering can be mathematically represented as follows~\cite{zeng2017energy}\revision{:}
\begin{equation}\label{eq:hovering}
    \mu_m(t) = 
        \underbrace{\frac{\delta}{8}\rho sA\Omega^3R^3}_{\textit{bladeprofile}, \,P_{o}}
        +\underbrace{\left(1+k\right)\frac{W^{3/2}}{\sqrt{2\rho A}}}_{\textit{induced}, \,P_{i}},\qquad \textit{(hovering)},
\end{equation}
where $\delta$, $\rho$, $s$, $A$, $\Omega$, $R$, $k$, and $W$ stand for the profile drag coefficient, air density, rotor solidity, rotor disc area, blade angular velocity, rotor radius, an incremental correction factor to induced power, and aircraft weight including battery and propellers, respectively. 
In addition, the energy expenditure model of \revision{the} $m$-th UAV during round-trip traveling can be mathematically expressed as follows~\cite{zeng2017energy}\revision{:}
\begin{equation}\label{eq:roundtrip}
\begin{split}
    \mu_m(t) = 
        \underbrace{P_{o}\left(1+\frac{3v^{2}}{U_{tip}}\right)}_{\textit{bladeprofile}}
        +\underbrace{P_{i}\left(\sqrt{1+\frac{v^{4}}{4v_{0}^{4}}}-\frac{v^{2}}{2v_{0}^{2}}\right)^{0.5}}_{\textit{induced}}\\
        +\underbrace{\frac{1}{2}d_{0}\rho sAv^{3}}_{\textit{parasite}},\qquad\textit{(round-trip traveling)},
\end{split}
\end{equation}
where $P_o$, $P_i$, $v$, $U_{tip}$, $v_0$, and $d_0$ denote with the blade profile power, induced power, flight speed, tip speed of the rotor blade, mean rotor-induced velocity, and fuselage drag ratio, respectively.

\subsection{Noise Distribution Model in UAV Positioning}\label{sec:3-c}

\begin{figure}[ht]
    \centering
    \subfigure[State Noise.]{
    \includegraphics[width=0.465\linewidth]{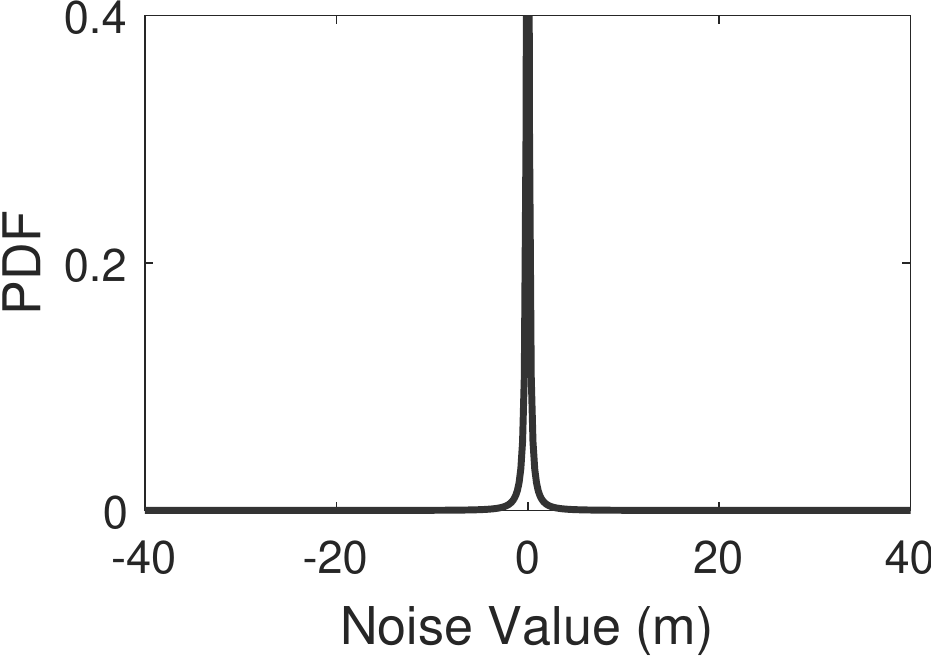}
    \label{fig:state_noise}
    }
    \subfigure[Action Noise.]{
    \includegraphics[width=0.465\linewidth]{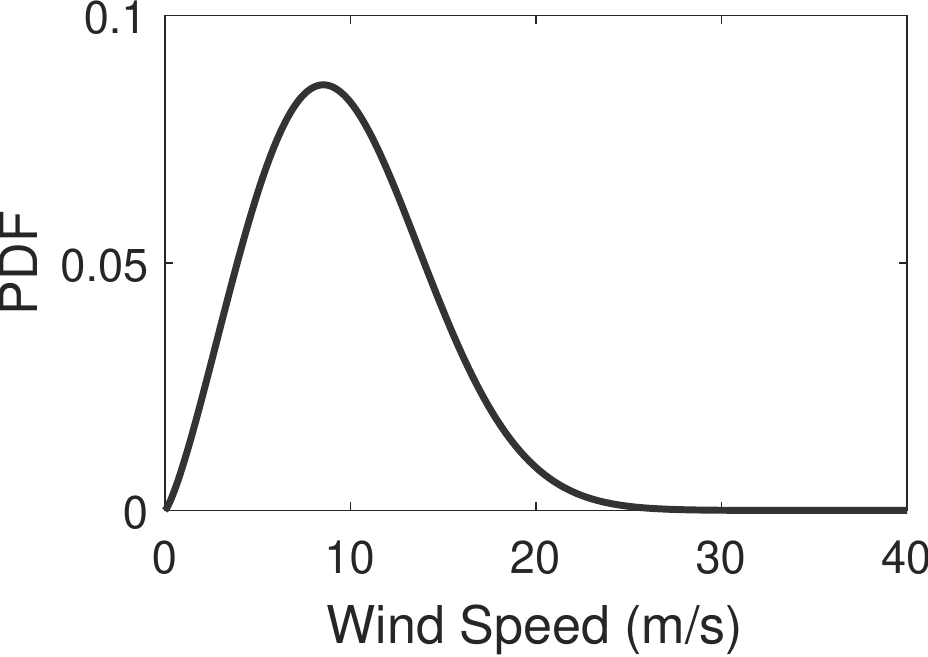}
    \label{fig:action_noise}
    }\\
    \subfigure[Wind Direction.]{
    \includegraphics[width=\linewidth]{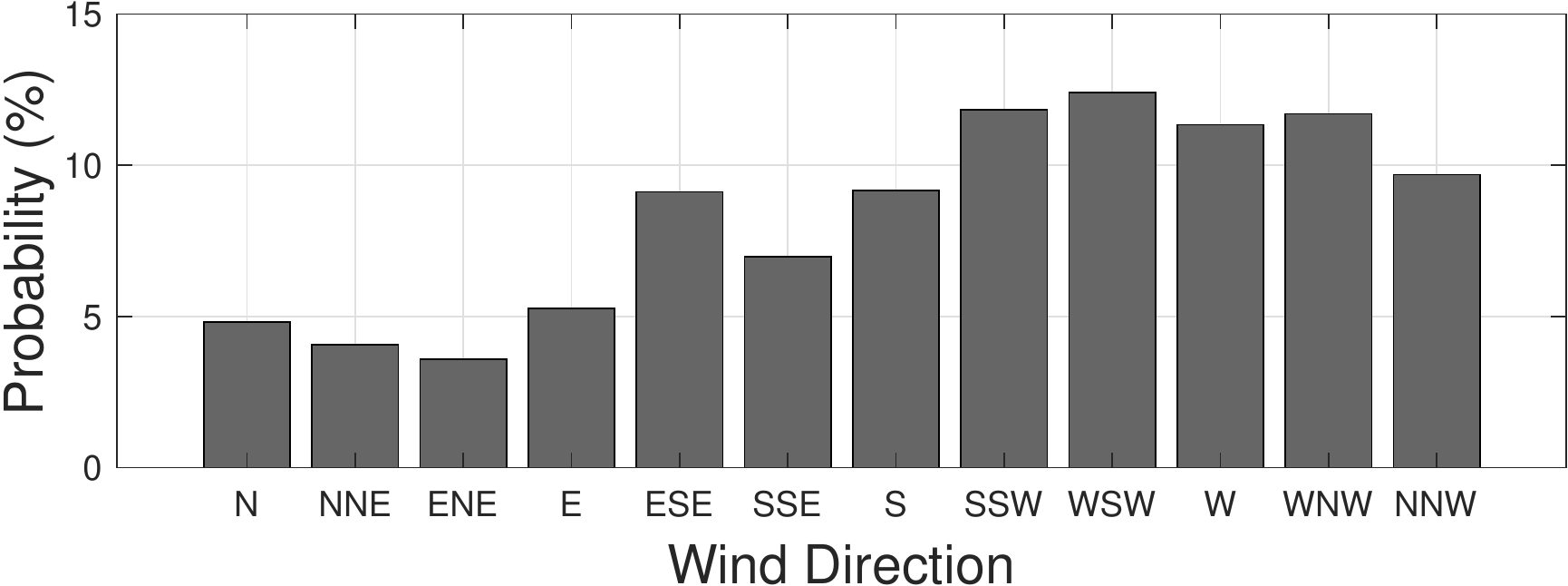}
    \label{fig:wind_direction}
    }
    \caption{Probability distribution functions of state and action noises.}
    \label{fig:Noise_model}
\end{figure}

\BfPara{State Noise}
UAVs trained by {QMACN} determine their locations using a global positioning system (GPS) which is commonly accessible in many mobility systems and platforms~\cite{mobisys2010paek}. 
However, \revision{GPS sensors} have noise, \textit{i.e.}, interference, jamming, and delay lock loop (DLL), which are modeled as Gaussian or non-Gaussian distributions. A widely used model \textit{i.e.}, generalized Cauchy noise~\cite{liu2008performance} is used to define the state noise in GPS sensors as shown below.
\begin{equation}
    p_{cy}(z,\sigma_z)=\frac{Y}{\left(1+v^{-1}[|z|/X]^k\right)^{v+1/k}},
    \label{eq:state_noise}
\end{equation}
where $X=\Big[\sigma_z^2\frac{\Gamma(1/k)}{\Gamma(3/k)}\Big]^{1/2}, Y=\frac{kv^{-1/k}\Gamma(v+1/k)}{2X\Gamma(v)\Gamma(1/k)}$, and $\Gamma(t)=\int_{0}^{\infty} x^{t-1}e^{-x}\, dx$ (Gamma function). In the variance of the generalized Cauchy density, $k$ and $v$ mean the impulsiveness and variance of the GPS noise~\cite{kassam2012signal}. We depict the probability distribution function (PDF) of the state noise as a generalized Cauchy density in Fig.~\ref{fig:state_noise} with the parameter $\sigma_z$ where we set $k=0.20$, $v=40$, and $\sigma_z^2=0.22$~\cite{liu2008performance}.

\BfPara{Action Noise}
The noise is modeled with \revision{a} Weibull distribution, a widely used probability distribution for \revision{modeling} wind speed\revision{. As per~\cite{feng2015modelling}, it is represented as follows:}
\begin{equation}
    p_{wb}(v,\sigma_v)=\left(c/A\right)\left(v/A\right)^{c-1}\texttt{exp}(-vc/A),
    \label{eq:action_noise}
\end{equation}
where $v$ is a wind velocity, $A=\left(\frac{\sigma_v}{\Bar{v}}\right)^{-1.086}$ and $c=\frac{\Bar{v}}{\Gamma(1+1/A)}$ stand for scale and shape parameters with the mean $\bar{v}$ and the standard deviation $\sigma_v$ of wind speed calculated from the data~\cite{feng2015modelling}. The meteorological measurements for a total of $145,048$ data were conducted during three years (from June 1st, 1999 to May 31, 2002). Fig.~\ref{fig:action_noise} is the PDF of Weibull distribution with $A=10.97\,m/s$ and $c=2.29$ to represent the probability of the wind speed. In addition, Fig.~\ref{fig:wind_direction} means the probability of $12$ directions where the wind can blow from.
Note that \textsf{N} refers to north, \textsf{S} refers to south, \textsf{E} refers to east and \textsf{W} refers to west direction respectively.

\BfPara{Modeling the Gap between Real-World and Ideal}
Considering the aforementioned noise states, we reconfigure the \textit{actor-critic} model. This paper considers both \textit{ideal} and \textit{real} state/action cases. The ideal case does not consider the noise in the environment, and the other considers the noise. The relationship between these cases can be modeled as follows\revision{:}
\begin{align}
    s_{\textrm{real}} & = s_{\textrm{ideal}} + n_s,
    \\
    a_{\textrm{real}} & = a_{\textrm{ideal}} + n_a,
\end{align}
where $n_s$ and $n_a$ stand for the noises of states and actions. Note that $n_s\sim p_{cy}(z, \sigma_z)\;\{x,y\}\in z$ and $n_a \sim p_{wb}(v, \sigma_v)$, where $x$ and $y$ are the cardinal points in the two-dimensional coordinate of UAVs' positions. 

\section{MARL Algorithm for Multi-UAV Cooperation}\label{sec:5}
\subsection{MARL Formulation}\label{sec:5-A}


Our considered multi-UAV system is mathematically defined as the decentralized partially observable MDP (Dec-POMDP), which is widely used for CTDE MARL framework~\cite{rashid2020monotonic}, because each UAV conducts distributed sequential decision-making with partial environment information due to its physical limitations.
The Dec-POMDP of $M$ UAVs \revision{is} denoted as $\langle\mathcal{S}, \mathcal{A}, P, r, \mathcal{O}, \mathcal{Z}, \gamma\rangle$ where 
    $\mathcal{S}$ is a set of states where $s\in\mathcal{S}$; 
    $\mathcal{A}$ is a set of actions where ${a_m}\in A_m \subset \mathcal{A}$, and here, $a_m$ denotes the $m$-th UAV's action, which composes the joint UAV actions, \textit{i.e.}, $\textbf{a}\in \mathcal{A}$, and note that $A_m$ means a set of \revision{the} $m$-th UAV's actions; 
    $P$ is a state transition probability $P:\Pr(s'\,|\,s,\textbf{a})=\mathcal{S}\times\mathcal{A}\times\mathcal{S}\rightarrow\mathcal{S}$ with joint UAV actions $\textbf{a}$; 
    $r$ is a reward which is given to every UAV with $r(s,\textbf{a},s')=\mathcal{S}\times\mathcal{A}\times\mathcal{S}\rightarrow\mathbb{R}$.
    In addition, 
    $\mathcal{O}$ is a set of observations where \revision{the} $m$-th UAV's observation is denoted as $o_m\in O_m \subset \mathcal{O}$ and note that $O_m$ means a set of observations of \revision{the} $m$-th UAV. Here, the joint UAV observation for training is denoted as $\textbf{o}\in\mathcal{O}$.
    $\mathcal{Z}$ is conditional observation probabilities function space, \textit{i.e.}, $\mathcal{Z}(s',\textbf{a},\textbf{o})$ and $\gamma$ is a discount factor.

When each $m$-th UAV takes action $a_m$ while observing $\mathcal{O}_m$ based on $\mathcal{Z}(s',\textbf{a}, \textbf{o})=\mathcal{S}\times\mathcal{A}\rightarrow\mathcal{O}$, the state is updated based on $P$. After this computation, the reward is \revision{generated based on} $r(s,\textbf{a},s')$. More details about observations, states, actions, rewards, and objective \revision{are described in the following.}

\BfPara{Observations}
The $m$-th UAV partially observes its own environment information based on its position \revision{$p_m\in\{x_m,y_m\}$ where $x_m$ and $y_m$ stand for Cartesian coordinates.} \revision{The $m$-th} UAV observes the distance between its own position and other UAV within \revision{its} observable scope, \textit{i.e.}, 
\begin{equation}
    d_{mm'} = \begin{cases}
        \|p_m - p_{m'}\|_2,& \textit{if.}~~~\|p_m - p_{m'}\|_2 \leq D_{\textit{th}}, \\
        -1,& \textit{(otherwise)},
    \end{cases}
\end{equation}
where $D_{\textit{th}}$ and $\|\cdot\|_2$ mean the observation scope and L2-norm\revision{, respectively}. 
\revision{In addition, it is necessary for every UAV to monitor its own energy level $e_m$ to avoid any malfunction that may result from the discharge of the battery.}
The observation of the $m$-th UAV is defined as \revision{$\mathcal{O}_m \triangleq\{p_m, \bigcup_{m'=1}^{M}\{d_{mm'}\}, e_m\}$}, \revision{$\forall m\in M$ and $\forall m'\in M$}.

\BfPara{States} 
The state consists of two service details\revision{:} \textit{i)} availability of service $c_{mn}\triangleq\{0,1\}$ and \textit{ii)} quality of service (QoS) $q_{mn}$. 
Here, $c_{mn}$ is a variable indicating whether \revision{the} $n$-th \revision{user} is serviced by \revision{the} $m$-th UAV by differentiating zero or not. The total support rate in the proposed network can be represented as $\sum_{m=1}^{M}\sum_{n=1}^{N}c_{mn}$. However, the quality of wireless communication services received by \revision{user}s varies depending on the requested service. Thus, this paper adopts a quality function~\cite{jung2021infrastructure} to evaluate the total QoS with the maximum supportable data rate while assuming that \revision{user}s randomly request different levels of data rates, including video streaming, online gaming, or web surfing~\cite{lai2019demand}. The detailed methodology for calculating actual data rates is described in Sec.~\ref{sec:6-2}.
The QoS of \revision{the} $n$-th user supported by \revision{the} $m$-th UAV can be formulated as follows\revision{:}
\begin{equation}
    q_{mn}=
    \begin{cases}
        \left(1+\exp^{-w_{a}\left(\kappa_{mn}-w_{b}\right)}\right)^{-1},\;\textit{(video traffic)},\\
        \log\left(w_{c}\cdot\kappa_{mn}+w_{d}\right),\;\textit{(otherwise)},
    \end{cases}
\label{eq:quality}
\end{equation}
where $\kappa_{mn}$ is the $n$-th user's data rates serviced by the $m$-th UAV. In addition, the values of weight parameters are $w_a$\,=\,0.01, $w_b$\,=\,1024, $w_c$\,=\,1, and $w_d$\,=\,1~\cite{jung2021infrastructure}. 
Here, all UAVs have the same state information simultaneously because the total service is affected by all UAVs.
Accordingly, the state information is denoted as $s\triangleq \bigcup^{M}_{m=1}\bigcup^{N}_{n=1}\{c_{mn}, q_{mn}\}$.

\BfPara{Actions}
All UAVs take actions sequentially based on their policies at $t$. UAVs can move in four cardinal points because they are in the two-dimensional Cartesian coordinates $(x,y)\in\mathbb{R}^2$. Therefore, the action set that UAVs can take is $\mathrm{a}\triangleq\{x_m \pm (v_m \times t), y_m \pm (v_m \times t)\}_{m=1}^M$ where $v_m$ is UAV's velocity.

\BfPara{Rewards}
The reward $r(s,\textbf{a})$ is generated with the current state $s$ and all UAVs' selected actions $\textbf{a}$. Then, the reward function is as follows\revision{:}
\begin{equation}
    r(s,\textbf{a},s')=w_c\times\sum_{m=1}^{M}\sum_{n=1}^{N}\left(c_{mn} \times q_{mn}\right)\times\tau\times\mathbbm{1}(e_m),
    \label{eq:reward}
\end{equation}
where $w_c$ is a reward weight to make the learning process more stable; $\tau^t$ is all UAVs' overlapped rate at $t$ which has to be decreased to reduce interference among them; $\mathbbm{1}(\cdot)$ is an indicator function to differentiate whether it is zero or not. Therefore, it can be seen that UAVs try to maximize the ground users' supported rate and QoS within the \revision{limited} energy according to \revision{the} reward function.

\BfPara{Objective}
Our main objective in MARL is formulated as \revision{follows}\revision{:}
\begin{equation}
\begin{split}
    & \pi^*_{\boldsymbol{\theta}} = \\
    & \argmaxD_{\boldsymbol{\theta}}\mathbb{E}_{s_{\textrm{real}}\sim E,\,\mathrm{a}_{\textrm{real}}\sim\pi_{\boldsymbol{\theta}}}\left[\sum_{t=1}^T\gamma^{t-1}\!\cdot\!r\left(s_{\textrm{real}},\mathbf{a}_{\textrm{real}},s'_{\textrm{real}}\right)\right],
    \label{eq:goal}
\end{split}
\end{equation}
where $E$, $T$, and $\gamma \in [0,1)$ stand for the environment where UAVs exist, an episode length, and a discounted factor. 
By redefining the objective function in RL, UAVs will consider realistic environmental noise (\textit{i.e.,} $s_{\textrm{real}}$, $\mathbf{a}_{\textrm{real}}$, and $s'_{\textrm{real}}$) to make their policy more robust to various types of noises. More details about how UAVs achieve optimal decision-making with the reconfigured objective function are discussed in Sec.~\ref{sec:4B}.

\subsection{QMACN for Cooperative Multi-UAV Mobile Access}\label{sec:4B}
\BfPara{QMARL Design}
Sec.~\ref{sec:1} clearly states the scalability issue due to the limited number of qubits. However, in the \textit{actor-critic} training method~\cite{konda1999actor}, the number of qubits must also increase as the number of agents grows in MARL. This need causes quantum errors that inhibit system stability~\cite{shor1995scheme,yun2022quantum}. 
Accordingly, we propose a novel {QMACN} algorithm utilizing the CTDE architecture, the methodology based on the multi-agent \textit{actor-critic} RL framework~\cite{yun2022quantum, lowe2017multi}. In CTDE, there is one \textit{centralized critic} and multiple \textit{actor} networks where the number of actor networks is commensurate with the number of agents in Dec-POMDP~\cite{oliehoek2016concise} as observed in the server of Fig.~\ref{fig:sys_pipeline}. CTDE-based agents make the sequential decision dispersively and train their \textit{actor} networks corresponding to the policy by evaluating the value of \textit{centralized critic} network, which can be expressed as follows\revision{:}
\begin{align}
\!\!\!Q(o,a;\boldsymbol{\theta}) &=\beta_a \langle O_a \rangle_{o,\boldsymbol{\theta}}\!\!\!\!&=\beta_a\text{Tr}(U^{a\dagger}(o;\boldsymbol{\theta})M_a U^{a}(o;\boldsymbol{\theta}))
\label{obs:actor}\\
\!\!\!V(s;\boldsymbol{\phi}) &=\beta_c \langle O \rangle_{s,\boldsymbol{\phi}}\!\!\!\!&=\beta_c\text{Tr}(U^{c\dagger}(s;\boldsymbol{\phi})M_c U^{c}(s;\boldsymbol{\phi}))
\label{obs:critic}
\end{align}
where operators $\text{Tr}(\cdot)$, $U(\cdot)$, and $(\cdot)^\dagger$ represent trace operator, the unitary operation for qubit rotation, and the entanglement of multiple qubits and complex conjugate, respectively. When the quantum state is measured, the output (known as observable) exists between -1 and 1, \textit{i.e.}, $\forall \langle O\rangle\in [-1,1]$, we utilize hyper-parameters $(\beta_a,\beta_c)$ for \textit{actors} and \textit{critic} networks to be well-trained.
Note that $M_a$ and $M_c$ are Hermitian matrices. With Eqs.~\eqref{obs:actor}--\eqref{obs:critic}, and the hyper-parameters, the \textit{actor-critic} networks  approximate the value function.

\BfPara{Quantum Actor}
At every time step $t$, the $m$-th quantum \textit{actor} chooses the action with the most significant probability among the currently possible actions based on its state and observation information, which is represented as follows\revision{:}
\begin{equation}
    \mathrm{a_{m,real}} = \argmaxD_{\mathrm{a}}\pi_{\boldsymbol{\theta}_m}(\mathrm{a}_{\textrm{ideal}}|s_{\textrm{ideal}}+n_s,{o}_m)+n_a,
\end{equation}
subject to
\begin{align}
    & \pi_{\boldsymbol{\theta}_m}(\mathrm{a}_{\textrm{ideal}}|s_{\mathrm{ideal}}+n_s,{o}_m) \triangleq \textit{softmax}(Q(o,a;\boldsymbol{\theta}_m)),\\
    & \textit{softmax}(\mathbf{x}) \triangleq \left[\frac{e^{x_1}}{\sum_{i=1}^N e^{x_i}},\cdots,\frac{e^{x_N}}{\sum_{i=1}^N e^{x_i}}\right],
\end{align}
where the \textit{softmax}$(\cdot)$ is an activation function to normalize the inputs. By using it, we  extract all actions' probabilities of the \textit{actor} with the observable $\langle O_a\rangle_{o,\boldsymbol{\theta}}$ in Eq.~\eqref{obs:actor}.

\BfPara{Quantum Centralized Critic}
The CTDE has a \textit{centralized critic} responsible for valuing the current state with a state-value function as follows\revision{:}
\begin{multline}
        V_{\boldsymbol{\phi}}(s) \!=\! \langle O_c \rangle_{s, \boldsymbol{\phi}} \!\simeq \! \\ 
        \mathbb{E}_{s_{\textrm{real}}\sim E,\,\mathrm{a}_{\textrm{real}}\sim\pi_{\boldsymbol{\theta}}}\left[\sum_{t'=t}^{T} \gamma^{t'-t}\!\!\cdot r(s_{\textrm{real}},\!\mathbf{a}_{\textrm{real}},s'_{\textrm{real}})\right],
\end{multline}
where $s_t$ is the measured state at the current state at $t$. We  also use the \textit{critic} network's observable to evaluate the current state's value.

\subsubsection{Training and Inference}
As mentioned in Eq.~\eqref{eq:goal}, agents in MARL try to maximize the expected return. We utilize the congruent state-value function $V_\phi$ of the \textit{centralized critic} network to derive the gradients from maximizing the common goal. With the parameters of \textit{actor} and \textit{critic} networks, which correspond to $\boldsymbol{\theta}$ and $\boldsymbol{\phi}$, we  configure a multi-agent policy gradient (MAPG) based on the temporal difference \textit{actor-critic} model by \textit{Bellman optimality equation}, as follows\revision{:}
\begin{multline}
    \nabla_{\boldsymbol{\theta}}J(\boldsymbol{\theta}) = \\ \mathbb{E}_{s_{\textrm{real}},\mathrm{o} \sim E} \left[\sum\limits^{T}_{t=1}\sum\limits^{M}_{m=1} 
    \delta_{\boldsymbol{\phi}}^t\cdot\nabla_{\boldsymbol{\theta}_m}\log\pi_{\boldsymbol{\theta}_m}(\mathrm{a}_m^t|s^t, o_m^t)\right], 
    \label{eq:l_actor}
\end{multline}
and
\begin{equation}
    \nabla_{\boldsymbol{\phi}}\mathcal{L}(\boldsymbol{\phi}) = \sum^{T}_{t=1}\nabla_{\boldsymbol{\phi}}\left\|\delta_{\boldsymbol{\phi}}^t\right\|^2,
    \label{eq:l_critic}
\end{equation}
subject to
\begin{equation}
    \delta_{\boldsymbol{\phi}}^t = r\left(s^t,\mathbf{a}^t,s^{t+1}\right)+\gamma V_{\boldsymbol{\phi}}(s^{t+1})-V_{\boldsymbol{\phi}}(s^t).
    \label{eq:delta}
\end{equation}

Among the three equations presented above, Eq.~\eqref{eq:l_actor} is the objective function for \textit{actor} networks, and the neural network parameters used in the equation are updated to be maximized by gradient ascent as follows\revision{:}
\begin{equation}
    \boldsymbol{\theta}_m^{t+1} \approx \boldsymbol{\theta}_m^{t} + \alpha_{\mathrm{actor}}\times[\,\delta^t_{\boldsymbol{\phi}} \!\cdot\!\nabla_{\boldsymbol{\theta}}\log\pi_{\boldsymbol{\theta}_m}(\mathrm{a}_m^t\,|\,s^t, \mathrm{o}_m^t)\,],
\end{equation}
where $\alpha_{\mathrm{actor}}$ stands for a learning rate of \textit{actor} networks. Eq.~\eqref{eq:l_critic} corresponds to the \textit{centralized critic} network's loss function which should be minimized by gradient descent as follows\revision{:}
\begin{equation}
    \boldsymbol{\phi}^{t+1} \approx \boldsymbol{\phi}^{t} + \alpha_{\mathrm{critic}}\times[\,\delta^t_{\boldsymbol{\phi}} \cdot \nabla_{\boldsymbol{\phi}}V_{\boldsymbol{\phi}}(s^{t})\,],
\end{equation}
where $\alpha_{\mathrm{critic}}$ is \textit{centralized critic} network's learning rate.

We describe how to obtain loss gradients with quantum and classical computing. Hereafter, we denote an \textit{actor}-network and \textit{critic} network as $\boldsymbol{\theta}$ and $\boldsymbol{\phi}$ for mathematical amenability.
\revision{T}heir loss values \revision{are calculated} with the temporal difference error of \textit{centralized critic} $\delta_{\boldsymbol{\phi}}$ in Eq.~\eqref{eq:delta}, where the derivative of \textit{actor}/\textit{critic}'s $i$-th parameters is expressed as follows\revision{:}
\begin{eqnarray}
    \frac{\partial J(\boldsymbol{\theta})}{\partial \theta_i} = \frac{\partial J(\boldsymbol{\theta})}{\partial \pi_{\boldsymbol{\theta}}} \cdot \frac{\partial \pi_{\boldsymbol{\theta}}}{\partial \langle O \rangle_{o,\boldsymbol{\theta}}}\cdot \frac{{\partial \langle O \rangle_{o,\boldsymbol{\theta}}}}{\partial \theta_i},\label{eq:loss_actor_derivative} \\
    \frac{\partial\mathcal{L}(\boldsymbol{\phi})}{\partial \phi_i}  = \frac{\partial\mathcal{L}(\boldsymbol{\phi})}{\partial V_{\boldsymbol{\phi}}} \cdot \frac{\partial V_{\boldsymbol{\phi}}}{\partial \langle O \rangle_{s,\boldsymbol{\phi}}} \cdot \frac{{\partial \langle O \rangle_{s,\boldsymbol{\phi}}}}{\partial \phi_i}, \label{eq:loss_critic_derivative}
\end{eqnarray}
where the first and second derivatives of RHS in Eqs.~\eqref{eq:loss_actor_derivative}--\eqref{eq:loss_critic_derivative} can be calculated by classical computing. However, the latter derivative cannot be calculated because the quantum state is unknown before its measurement. 
Thus, we use the parameter-shift rule~\cite{crook19}. This allows us to bridge classical and quantum computing by multiplying the partial derivatives of classical and quantum computing. The parameter-shift rule is applied to Eqs.~\eqref{eq:l_actor}--\eqref{eq:l_critic}, where the derivative of \revision{the} $i$-th \textit{actor} parameter is obtained with zeroth derivative as follows\revision{:}
\begin{equation}
   \frac{{\partial \langle O \rangle_{o,\boldsymbol{\theta}}}}{\partial \theta_i} =\langle O \rangle_{o,\boldsymbol{\theta} + \frac{\pi}{2} \mathbf{e}_i } - \langle O \rangle_{o,\boldsymbol{\theta} - \frac{\pi}{2} \mathbf{e}_i },\label{eq:param-shift}
\end{equation}
where $\mathbf{e}_i$ correspond to the $i$-th basis of $\boldsymbol{\theta}$, respectively. Similarly, the LHS of Eq.~\eqref{eq:loss_critic_derivative} is obtained via Eq.~\eqref{eq:param-shift}.
Finally, we  calculate the gradient of the objective function as elaborated in Eqs.~\eqref{eq:l_actor}--\eqref{eq:l_critic}.
\subsubsection{Algorithm Pseudo-Code}
Details of the proposed CTDE-based training and inference procedure are explained in Algorithm~\ref{alg:CTDE} and the corresponding descriptions are as follows\revision{:}
\begin{enumerate}
    \item
    Initialize the parameters of \textit{actor} and \textit{centralized critic} networks, which are $\boldsymbol{\theta}$ and $\boldsymbol{\phi}$, respectively \textit{(line 1)}.
    \item 
    All agents learn their policy by repeating the below procedure \textit{(lines 2-19)} until all training epochs reach maximum epochs:
    (i) At the start of every epoch, initialize environments to set starting state $s_0$ \textit{(line 3)}.
    (ii) Each UAV selects an action based on its policy for every time step. By taking action, the environment is transited to the next time step ${s}^{t+1}$ \textit{(lines 5–8)}. Calculate the total reward $\mathbf{r}^t$ and transition pairs (\textit{i.e.,} experiences) by UAVs $\xi = \{s^t, \textbf{o}^t, \text{$\mathbf{r}^t$}, s^{t+1}, \textbf{o}^{t+1}\}$ are stored in the replay buffer $\mathcal{D}$ \textit{(lines 8-10)}.
    (iii) UAVs randomly sample the mini-batch from $\mathcal{D}$ for getting $V_{\boldsymbol{\phi}}$. By doing so, the learning performance improves by reducing the continuity of the data used for training~\cite{mnih2013playing}. Note that we start training networks of UAVs to prevent the direction of training from being biased to initial data \textit{(lines 12-18)}.
    (iv) Update parameters of the \textit{centralized critic} network $\boldsymbol{\phi}$ by gradient descent to the loss function in Eq.~\eqref{eq:l_critic} to reduce its value \textit{(lines 14-15)}.
    (v) Update parameters of all \textit{actor} networks $\boldsymbol{\theta}$ by gradient ascent to the objective function in Eq.~\eqref{eq:l_actor} in the direction of increasing its value evaluated by \textit{centralized critic} network \textit{(line 16)}.
    After completing all UAVs' policies training, they perform multiple inference processes on the environment \textit{(lines 20-28)}.
\end{enumerate}

\begin{algorithm}[t]
\small
    Initialize weights of the \textit{actor} and \textit{centralized critic} networks which are denoted as $\boldsymbol{\theta}$ and $\boldsymbol{\phi}$, $\forall m \in [1,M]$ \\
    \For{Epoch = 1, MaxEpoch}{
        $\triangleright$ \textbf{Initialize Multi-UAV Environments}, set $s_0$ \\
        \For{time step = 1, $T$}{
            \For{each UAV $m$}{
                $\triangleright$ Select the action $\mathrm{a}_m$ based on its policy $\pi_{\boldsymbol{\theta}_m}(\mathrm{a}_m^t\,|\,s_\mathrm{real}^t,o_m^t)$ at time step $t$ \\
            }
            $\triangleright$ $s^t \rightarrow s^{t+1}$, $\textbf{o}^t \rightarrow \textbf{o}^{t+1}$ with the reward $\mathbf{r}^t$

            $\triangleright$ Set $\xi= \{s^t, \textbf{o}^t, \textbf{a}^t, \text{$\mathbf{r}^t$}, s^{t+1}, \textbf{o}^{t+1}\}$ \\
            $\triangleright$ Update replay buffer $\mathcal{D}$: $Enqueue(\mathcal{D},\xi)$ \\
        }
        \If {$\mathcal{D}$ \textbf{is full enough to train:}}{
        
        \For{each UAV $m$}{
        $\triangleright$ Get $V_{\boldsymbol{\phi}}$ by sampling mini-batch $\mathcal{B}$ from $\mathcal{D}$\\
        
        $\triangleright$ Update $\boldsymbol{\phi}$ by \textbf{gradient descent} to loss function of the \textit{centralized critic} network: $\nabla_{\boldsymbol{\phi}}\mathcal{L}(\boldsymbol{\phi})$ \\
        
        $\triangleright$ Update $\boldsymbol{\theta}_m$ by \textbf{gradient ascent} to objective function of the \textit{actor} network: $\nabla_{\boldsymbol{\theta}_m}J(\boldsymbol{\theta}_m)$
        }
    }
    }
    \For{Episode = 1, MaxEpisode}{
        $\triangleright$ \textbf{Initialize Multi-UAV Environments}, set $s_0$ \\
        \For{time step = 1, $T$}{
            \For{each UAV $m$}{
                    $\triangleright$ Select the action $\mathrm{a}_m$ based on its policy $\pi_{\boldsymbol{\theta}_m}(\mathrm{a}_m^t\,|\,s_\mathrm{real}^t,o_m^t)$ at time step $t$ \\
                }
                $\triangleright$ $s^t \rightarrow s^{t+1}$, $\textbf{o}^t \rightarrow \textbf{o}^{t+1}$ with the reward $\mathbf{r}^t$ \\
        }
    }
    \caption{QMACN training and inference process for multi-UAV cooperation}
    \label{alg:CTDE}
\end{algorithm}

\begin{table}[t]
\centering
\caption{UAV Parameter Specification~\revision{\cite{tvt202106jung}}.}
\renewcommand{\arraystretch}{1.0}
\scriptsize
\begin{tabular}{c||c}
\toprule[1pt]
\textbf{Notation} & \textbf{Value} \\ \midrule
Flight speed, $v$ & 20\,$[m/s]$ \\
Average maximum flight time & 30\,$[min]$ \\
Capacity of flight battery & 5,870\,$[mAh]$\\
Voltage & 15.2\,$[V]$\\
Aircraft weight including battery and propellers, $W$ & 1375\,$[g]$\\ Rotor radius, $R$ & 0.4\,$[m]$ \\
Rotor disc area, $A=\pi R^{2}$ & 0.503\,$[m^{2}]$ \\
Number of blades , $b$ & $4$ \\
Rotor solidity, $s, \frac{0.0157b}{\pi R}$ & $0.05$ \\
Blade angular velocity, $\Omega$ & 300\,$[radius/s]$\\
Tip speed of the rotor blade , $U_{tip}=\Omega R^{2}$ & $120$ \\
Fuselage drag ratio, $d_{0}=\frac{0.0151}{sA}$ & $0.6$ \\
Air density, $\rho$ & 1.225\,$[kg/m^{3}]$ \\
Mean rotor-induced velocity in hovering, $v_{0}=\sqrt\frac{W}{s\rho A}$ & $4.03$ \\
Profile drag coefficient, $\delta$ & $0.012$ \\
Incremental correction factor to induced power, $k$ & $0.1$ \\
Maximum service ceiling & 6000\,$m$ \\
\bottomrule[1pt]
\end{tabular}
\label{tab:parameters of uav}
\end{table}

\begin{table}[t!]
\small
\caption{Experimental setup parameters}
\centering
\begin{tabular}{c|l|r}\toprule[1pt]
    \multicolumn{2}{c}{\textbf{{Parameters}}} & \textbf{{Values}} \\
    \midrule
    $M$ & The number of UAVs & $4$ \\
    $N$ & The number of \revision{user}s  & $25$ \\
    $T$ & Episode length & $30\,\mathrm{min}$ \\
    {$|\mathcal{S}|$} & State dimension & $179$ \\
    {$|\mathcal{A}|$} & Action dimension & $5$ \\
    {$|\mathcal{D}|$} & Replay buffer size & $50$k \\
    {$|\mathcal{B}|$} & Mini-batch size & $32$ \\
    $\gamma$ & Discount factor & $0.98$ \\
    $w_c$ & Reward coefficient & $0.01$ \\
    $\epsilon_\mathrm{init}$ & Initial epsilon & $0.275$ \\
    - & Annealing epsilon & $0.00005$ \\
    $\epsilon_\mathrm{min}$ & Minimum epsilon & $0.01$ \\
    - & Training epochs & $10$k \\
    {$\alpha_{\mathrm{actor}}$} & Learning rate of \textit{actor} & $0.001$ \\
    {$\alpha_{\mathrm{critic}}$} & Learning rate of \textit{centralized critic} & $0.00025$ \\
    - & Activation function & ReLU \\
    - & Optimizer & Adam \\
    \bottomrule[1pt]
\end{tabular}
\label{tab:parameter}
\end{table}

\begin{table}[t]
\footnotesize
\caption{Computing Platform Specifications}
\label{tab:platform}
\begin{center}
	\centering
	\begin{tabular}{l|r}
    \toprule[1.0pt]
    \centering
      \text{\textbf{System}} & \text{\textbf{Specification}} \\
    \midrule[1.0pt]
    CPU & Intel(R) Core(TM) i7-9700K CPU @3.60 GHz \\
        & RAM: 64.0 GB \\ 
    \midrule
    GPU & NVIDIA GeForce RTX 3090 \\ 
        & The number of cores: $10,496$ \\
        & Memory: 24 GB GDDR6X\\
    \midrule
    Platform  & Memory: DDR4 64 GB \\
    (PC)      & SSD: 500 GB (NVMe) \\
              & HDD: 1 TB \\
              & VGA: RTX 3090 \\
    \midrule
    Software  & Conda: v\,22.11.1 \\
    Library   & Python: v\,3.9.15 \\
              & Pytorch: v\,1.13.1 \\
              & Torchquantum: v\,0.1.0 \\
              & Numpy: v\,1.23.4 \\
    \bottomrule[1.0pt]
	\end{tabular}
\end{center}
\end{table}

\revision{
\subsection{Computational Complexity Analysis}\label{sec:5-c}
In this section, the complexity analysis of QNN is elaborated\footnote{Hereafter, we use the operator $| \cdot |$ to represent the dimension of \revision{a} vector.}.
First, the number of operations required for the execution of the quantum actor will be calculated.  
Suppose that the input dimension is $|\mathcal{O}|$, the output dimension is $|\mathcal{A}|$ and the number of parameters is $|\boldsymbol{\theta}|$.
Then, the number of operations in the state encoder will be proportional to the number of input variables, \textit{i.e.}, $|\mathcal{O}|$.
Furthermore, the number of parameterized gate operations will be proportional to the number of gates as well, \textit{i.e.}, $|\boldsymbol{\theta}|$. 
In addition, $|\mathcal{A}|$ quantum measurements and softmax operations are required to produce action decisions.
Finally, by considering all the operations, the computational complexity of quantum actor execution is summarized as $\mathcal{C}_{QA} = \Omega(|\mathcal{O}| + |\boldsymbol{\theta}| +  |\mathcal{A}|)$.
Similarly, the computational complexity of quantum centralized critic is calculated by replacing the input dimension, output dimension, and the number of parameters to $|\mathcal{S}|$,  1, and $|\boldsymbol{\phi}|$, respectively. Then, we have the computational complexity of quantum centralized critic as $\mathcal{C}^E_{QC} = \Omega(|\mathcal{S}| + |\boldsymbol{\phi}| +  1)$.

Next, we will elaborate on the computing cost for training cost per epoch.
To calculate the target difference in Eq.~\eqref{eq:delta}, $2 \cdot T \cdot \mathcal{C}^E_{QC}$ number of operations are needed. 
The latter term in the right hand-side of Eq.~\eqref{eq:l_actor} can be obtained with $ 2 \cdot T \cdot M \cdot (|\mathcal{A}| + \mathcal{C}^E_{QA} )$ operations.
In summary, the training cost of QMACN per epoch can be represented as $\mathcal{C}_{QMACN} = \Omega(T\cdot( \mathcal{C}^E_{QC}+  M \cdot (|\mathcal{A}|+C^E_{QA})))$.
In contrast to QMACN, the classical multi-agent actor-critic has a computational complexity of  
$\mathcal{C}^E_{CA}= \Omega(|\mathcal{O}| \!\cdot \!|\boldsymbol{\theta}|\! \cdot\! |\mathcal{A}|)$ for execution of actor, 
$\mathcal{C}^E_{CC}= \Omega(|\mathcal{S}| \!\cdot \!|\boldsymbol{\phi}|)$ for execution of critic, and 
$\mathcal{C}_{MACN}= \Omega(T \!\cdot\! (M \!\cdot\! |\mathcal{O}|\! \cdot \!|\boldsymbol{\theta}| \!\cdot \!|\mathcal{A}| + |\mathcal{S}| \!\cdot\! |\boldsymbol{\phi}|))$ for training cost, respectively. These equations indicate that the computation complexity of QMACN is lower than the complexity of classical multi-agent actor-critic.
}

\section{Performance Evaluation}\label{sec:6}
\subsection{System Setup}
\subsubsection{Environment}\label{sec:6-1}
Our environment setting includes two primary objects, ground users and UAVs. \revision{User}s are randomly distributed over $6\,km\times6\,km$ area/map; and UAVs fly at \revision{an} altitude of $2500\,\mathrm{m}$. We assume that all UAVs are located at the center of the map at the start of every episode. 
Throughout the training phase, an $\epsilon$-greedy strategy is utilized \revision{to control exploration and exploitation} of UAVs.
In addition, weights of neural networks are updated with the Adam optimizer. 
\revision{To achieve a practical representation of UAV mechanics, this paper employs the \textit{DJI Phantom4 Pro v2.0} model~\cite{tvt202106jung}.
The energy-associated parameters for the selected UAV model, as presented in Eqs.~\eqref{eq:hovering}--\eqref{eq:roundtrip}, are compiled in Table~\ref{tab:parameters of uav}\cite{tvt202106jung}}.
Specifications of other experimental parameters and computing platforms are summarized in Tables~\ref{tab:parameter}--\ref{tab:platform}. Lastly, notice that the learning parameters and hyperparameters used in our model were obtained after exhaustive experiments trying to optimize performance, and these experiments are omitted for brevity.

\begin{figure*}[!ht]
    \centering
    \includegraphics[width=0.9\linewidth]{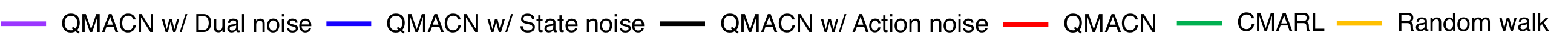}\\
    \subfigure[Average Total Reward.]{
    \includegraphics[width=0.31\linewidth]{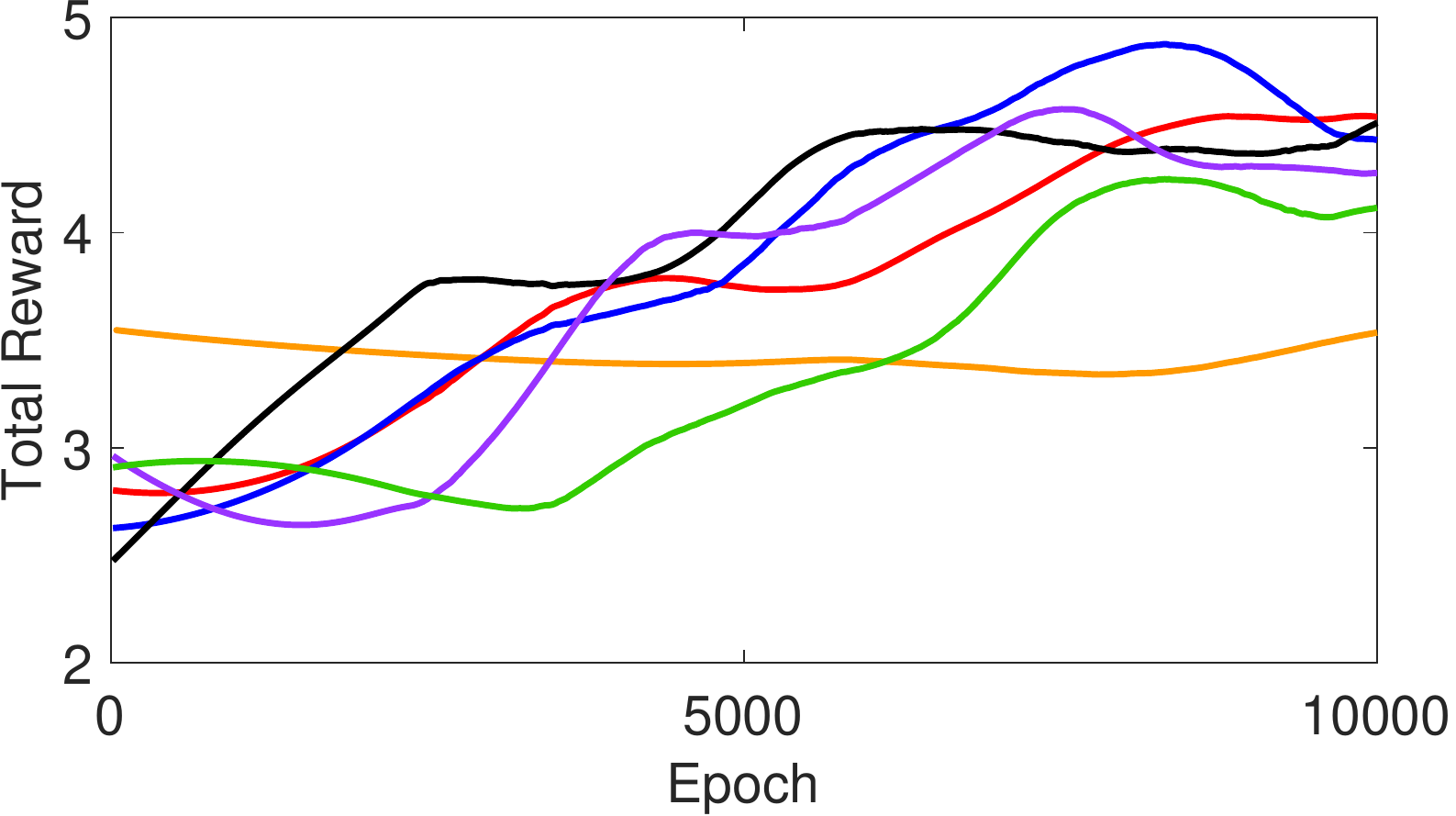}
    \label{fig:reward}
    }
    \subfigure[Average Support Rate.]{
    \includegraphics[width=0.31\linewidth]{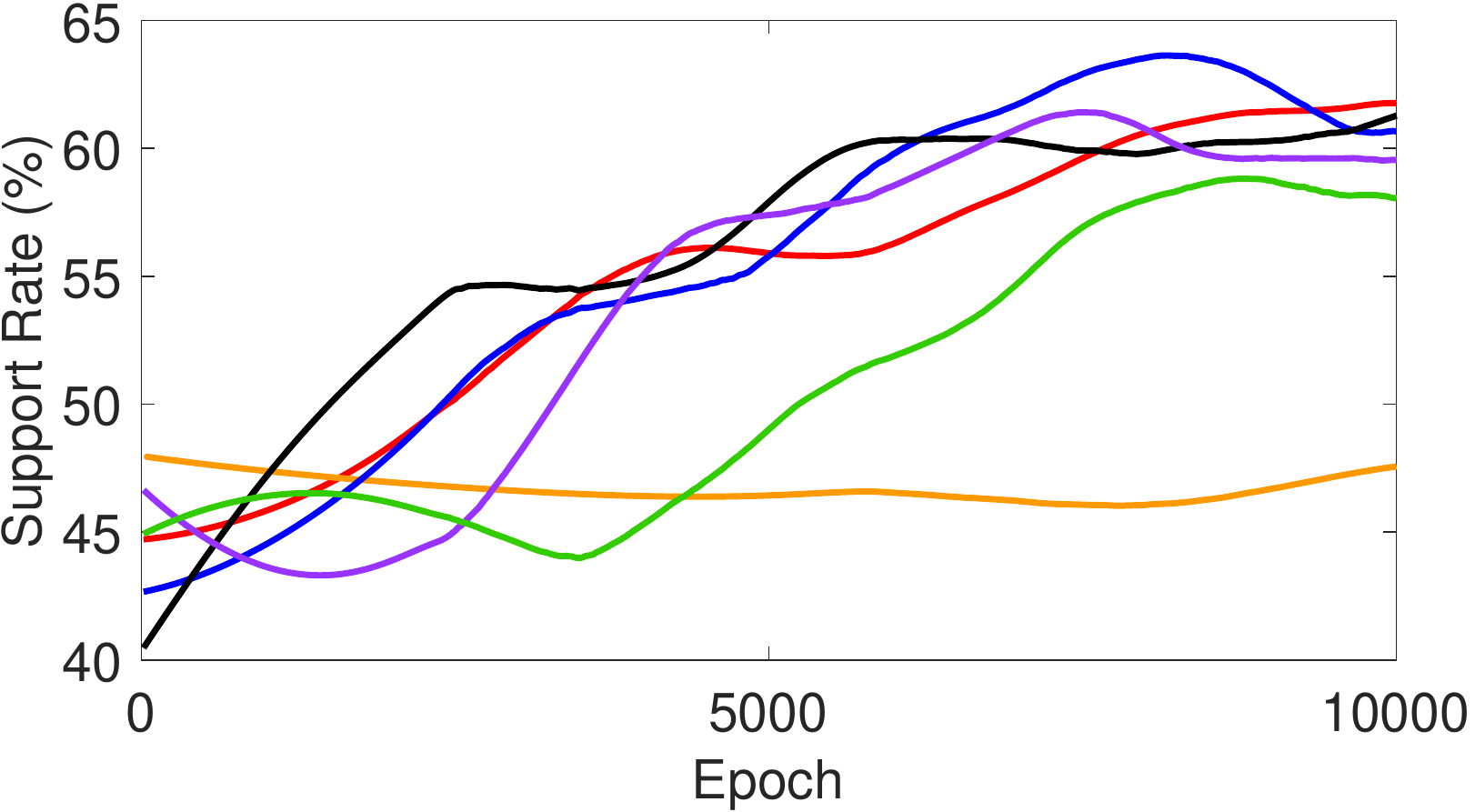}
    \label{fig:support_rate}
    }
    \subfigure[Average Total QoS.]{
    \includegraphics[width=0.31\linewidth]{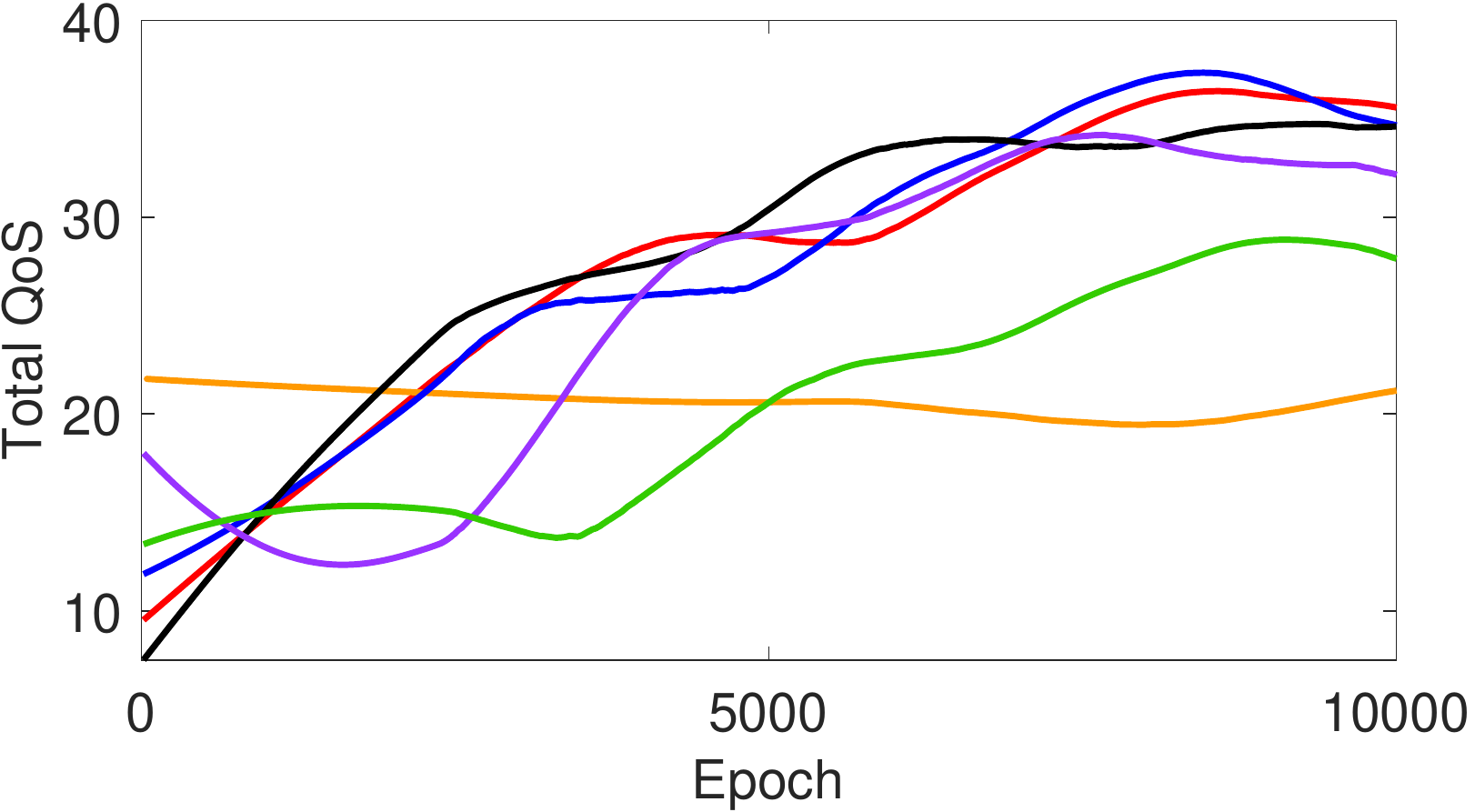}
    \label{fig:qos}
    }
    \caption{Average values of training results in all training benchmarks in every time step. Fig.~\ref{fig:reward} shows UAVs' total obtained rewards over the entire epochs, Fig.~\ref{fig:support_rate} shows UAVs' total obtained average support rate over the entire epochs, and Fig.~\ref{fig:qos} shows the ground \revision{user}s' support rate and QoS served by UAVs.}
    \label{fig:overall_result}
\end{figure*}

\begin{table*}[t!]
    \centering
    \caption{Relative performance of QMACN-based benchmarks \\ over the classical neural network-based CMARL (unit: \%)}
    \resizebox{1.13\columnwidth}{!}{\begin{minipage}[h]{1.13\columnwidth}
    \centering
    \label{tab:training}
    \newcolumntype{R}{>{\raggedleft\arraybackslash}X}
\begin{tabularx}{1\linewidth}{c l l l}
    \toprule[1pt]
    \multicolumn{4}{c}{\textbf{\circled{1} : Relative Training Superiority of QMARL algorithms than CMARL}} \\
    \midrule[.5pt]
    \ \ \textit{Benchmark} & \ \ \ \ \ \ Fig.\ref{fig:reward} & \ \ \ \ \ \ Fig.\ref{fig:support_rate} & \ \ \ \ \ \ Fig.\ref{fig:qos} \\     
    \cmidrule(lr){1-1} \cmidrule(lr){2-2} \cmidrule(lr){3-3} \cmidrule(lr){4-4}
    \textbf{QMACN} & {$21.27\,\%$} \hspace{1pt}  \tikz{
        \fill[fill=color1] (0.0,0) rectangle (1,0.2);
        \fill[pattern=north west lines, pattern color=black!30!color1] (0.0,0) rectangle (1,0.2);
    
    } & {$27.44\,\%$} \hspace{1pt}  \tikz{
        \fill[fill=color1] (0.0,0) rectangle (1,0.2);
        \fill[pattern=north west lines, pattern color=black!30!color1] (0.0,0) rectangle (1,0.2);
    } & {$6.390\,\%$} \hspace{1pt}  \tikz{
        \fill[fill=color1] (0.0,0) rectangle (1,0.2);
        \fill[pattern=north west lines, pattern color=black!30!color1] (0.0,0) rectangle (1,0.2);
    } \\
    w/\,State\,noise & {$15.98\,\%$} \hspace{1pt}  \tikz{
        \fill[fill=color2] (0.0,0) rectangle (0.7513,0.2);
        \fill[pattern=north west lines, pattern color=black!30!color2] (0.0,0) rectangle (0.7513,0.2);
    } & {$24.22\,\%$} \hspace{1pt}  \tikz{
        \fill[fill=color2] (0.0,0) rectangle (0.8827,0.2);
        \fill[pattern=north west lines, pattern color=black!30!color2] (0.0,0) rectangle (0.8827,0.2);
    } & {$4.513\,\%$} \hspace{1pt}  \tikz{
        \fill[fill=color2] (0.0,0) rectangle (0.7063,0.2);
        \fill[pattern=north west lines, pattern color=black!30!color2] (0.0,0) rectangle (0.7063,0.2);
    } \\
    w/\,Action\,noise & {$19.36\,\%$} \hspace{1pt}  \tikz{
        \fill[fill=color8] (0.0,0) rectangle (0.91,0.2);
        \fill[pattern=north west lines, pattern color=black!30!color3] (0.0,0) rectangle (0.91,0.2);
    } & {$23.83\,\%$} \hspace{1pt}  \tikz{
        \fill[fill=color8] (0.0,0) rectangle (0.8684,0.2);
        \fill[pattern=north west lines, pattern color=black!30!color3] (0.0,0) rectangle (0.8684,0.2);
    } & {$5.460\,\%$} \hspace{1pt}  \tikz{
        \fill[fill=color8] (0.0,0) rectangle (0.8545,0.2);
        \fill[pattern=north west lines, pattern color=black!30!color3] (0.0,0) rectangle (0.8545,0.2);
    } \\
    w/\,Dual\,noise & {$8.150\,\%$} \hspace{1pt}  \tikz{
        \fill[fill=color4] (0.0,0) rectangle (0.3832,0.2);
        \fill[pattern=north west lines, pattern color=black!30!color4] (0.0,0) rectangle (0.3832,0.2);
    } & {$15.28\,\%$} \hspace{1pt}  \tikz{
        \fill[fill=color4] (0.0,0) rectangle (0.5569,0.2);
        \fill[pattern=north west lines, pattern color=black!30!color4] (0.0,0) rectangle (0.5569,0.2);
    } & {$2.566\,\%$} \hspace{1pt}  \tikz{
        \fill[fill=color4] (0.0,0) rectangle (0.4016,0.2);
        \fill[pattern=north west lines, pattern color=black!30!color4] (0.0,0) rectangle (0.4016,0.2);
    } \\
    \bottomrule[1pt]
\end{tabularx}
    \end{minipage}}
\end{table*}

\subsubsection{Communication Methodology}\label{sec:6-2}
In addition, a 60\,GHz mmWave wireless technology is considered for communications. 
The reason why the 60\,GHz wireless network is considered is that it has \textit{i)} a large channel bandwidth, \textit{ii)} low-latency transmission, \textit{iii)} high beam directivity, \textit{iv)} high diffraction, and \textit{v)} high scattering~\cite{singh2011interference}. 
These characteristics are advantageous due to less sensitivity to interference from nearby mobile access (spatial reuse) while being sensitive to blocking~\cite{singh2011interference}.
In an urban environment, for example, blocking by high-rise buildings may affect mmWave wireless systems. 
However, our proposed algorithm minimizes this problem through the optimal cooperative UAVs positioning. 
Thus, it is possible to provide better QoS to users using high directivity as well as low latency~\cite{singh2011interference}. \revision{More details regarding the communication channel in urban environments are in~\cite{park2022cooperative} and also summarized in Appendix~\ref{sec:mmwave}.}

\begin{table}[ht!]
\centering
\scriptsize
\caption{Comparative Analysis of QMACN}
\begin{tabular}{l||cccccc}
\toprule[1pt]
\textbf{Method}/\textbf{Specification} & QC & NN & State Noise & Action Noise \\
\midrule
{QMACN (w/ Dual Noise)} & \checkmarks & \checkmarks & \checkmarks & \checkmarks \\
{QMACN (w/ State Noise)} & \checkmarks & \checkmarks & \checkmarks & \crossmark \\
{QMACN (w/ Action Noise)} & \checkmarks & \checkmarks & \crossmark & \checkmarks \\
{QMACN (Ideal)} & \checkmarks & \checkmarks & \crossmark & \crossmark \\
{CMARL} & \crossmark & \checkmarks & \crossmark  & \crossmark \\
{Random Walk} & \crossmark & \crossmark & \crossmark  & \crossmark \\
\bottomrule[1pt]
\end{tabular}
\label{tab:comparison}
\vspace{-1mm}
\end{table}

\subsubsection{Benchmark}
We will compare the performance of our proposed method against various benchmarks to substantiate the proposed QMARL, as outlined below.
\begin{itemize}
    \item `\textbf{QMACN w/ Dual noise}' trains each UAV's policy with our QMARL in a realistic environment with state and action noise. Accordingly, the position of the UAV is affected by not only its action decision. We conduct an \textit{ablation study} to investigate how each noise component affects the performance and robustness of our model against noise.
    \item `\textbf{QMACN w/ State noise}' and `\textbf{QMACN w/ Action noise}' train the policy with QMARL with only state noise from GPS sensors and action noise from winds, respectively.
    \item `\textbf{{QMACN}}' trains the policy using QMARL in an environment without noise. It means that the position of the UAV is only dependent on its action decision.
    \item `\textbf{CMARL}' is the latest MARL using classical neural networks to compare the training performance with our proposed QMARL. The multi-UAVs in this training scheme train their policies by the classical CTDE using backpropagation~\cite{hecht1992theory}. Note that we set the same number of model parameters with our proposed QMARL training benchmarks.
    \item `\textbf{Random walk}' algorithm means all UAVs decide their actions randomly regardless of given state and observation information. By comparing it, we verify the superiority of MARL.
\end{itemize}

\subsection{Feasibility of Reward Function Design}
In this subsection, we will scrutinize the feasibility of reward function design with Fig.~\ref{fig:overall_result}. 
From the results, we observe that noise-reflected benchmarks denoted as `QMACN w/ Dual noise', `QMACN w/ State noise', and `QMACN w/ Action noise' respectively show huge fluctuations such as high variance in data than the ideal model due to the environmental noise effect. 
However, the total reward of QMACN-based training algorithms converges to a more considerable value than 'CMARL' and 'Random walk' at the end of the training process. 
Despite fewer parameters, we emphasize that our proposed QMACN has a more predominant policy training performance than the CMARL as shown in Table~\ref{tab:training}.
UAVs trained via QMACN get analogous final reward values in the order of `QMACN', `QMACN w/ Action noise', `QMACN w/ State noise', and `QMACN w/ Dual noise'.
Next, the average support rate and QoS of ground \revision{user}s' tend to be similar to the convergence as observed in Figs.~\ref{fig:support_rate}--\ref{fig:qos}. These results show that our formulation about the reward function is well set for multi-UAVs to achieve the intended goals of reliable wireless communication service provisioning.
In summary, we confirm that our reward design is justified for not only ideal condition but noisy environment. 

\begin{figure}
    \centering
    \includegraphics[width=\linewidth]{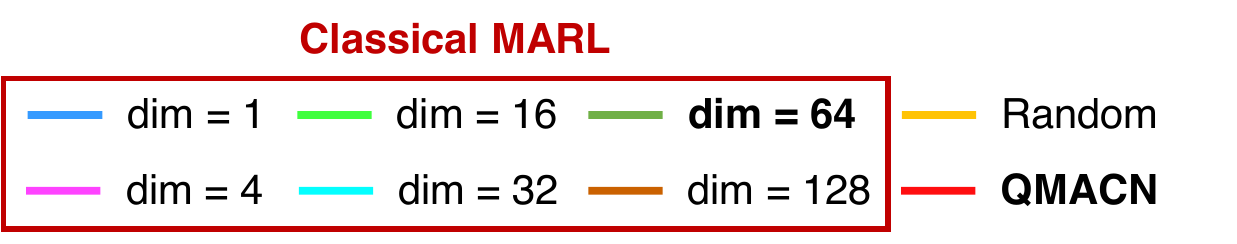}\\
    \includegraphics[width=0.7\linewidth]{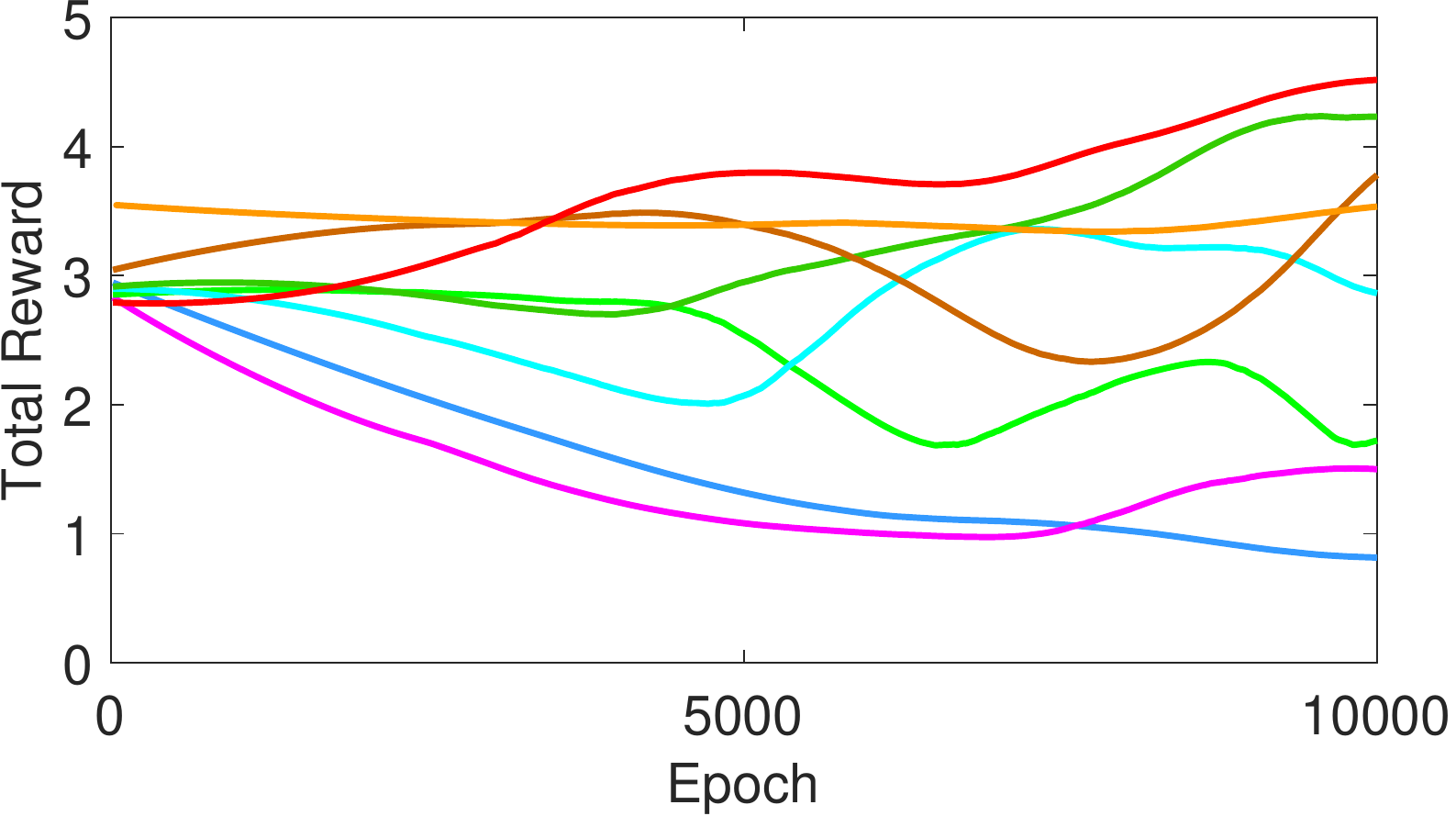}
    \caption{Reward convergence of classical neural networks and quantum computing in a noise-free (\textit{i.e.,} ideal) environment. \textit{dim} means the hidden layer dimension of each classical neural network.}
    \label{fig:with_classical_NN}
    \vspace{-5mm}
\end{figure}

\begin{figure}[t!]
    \centering
    \includegraphics[width=0.99\linewidth]{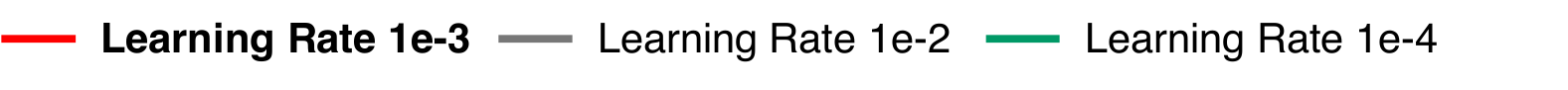}\\
    \includegraphics[width=0.7\linewidth]{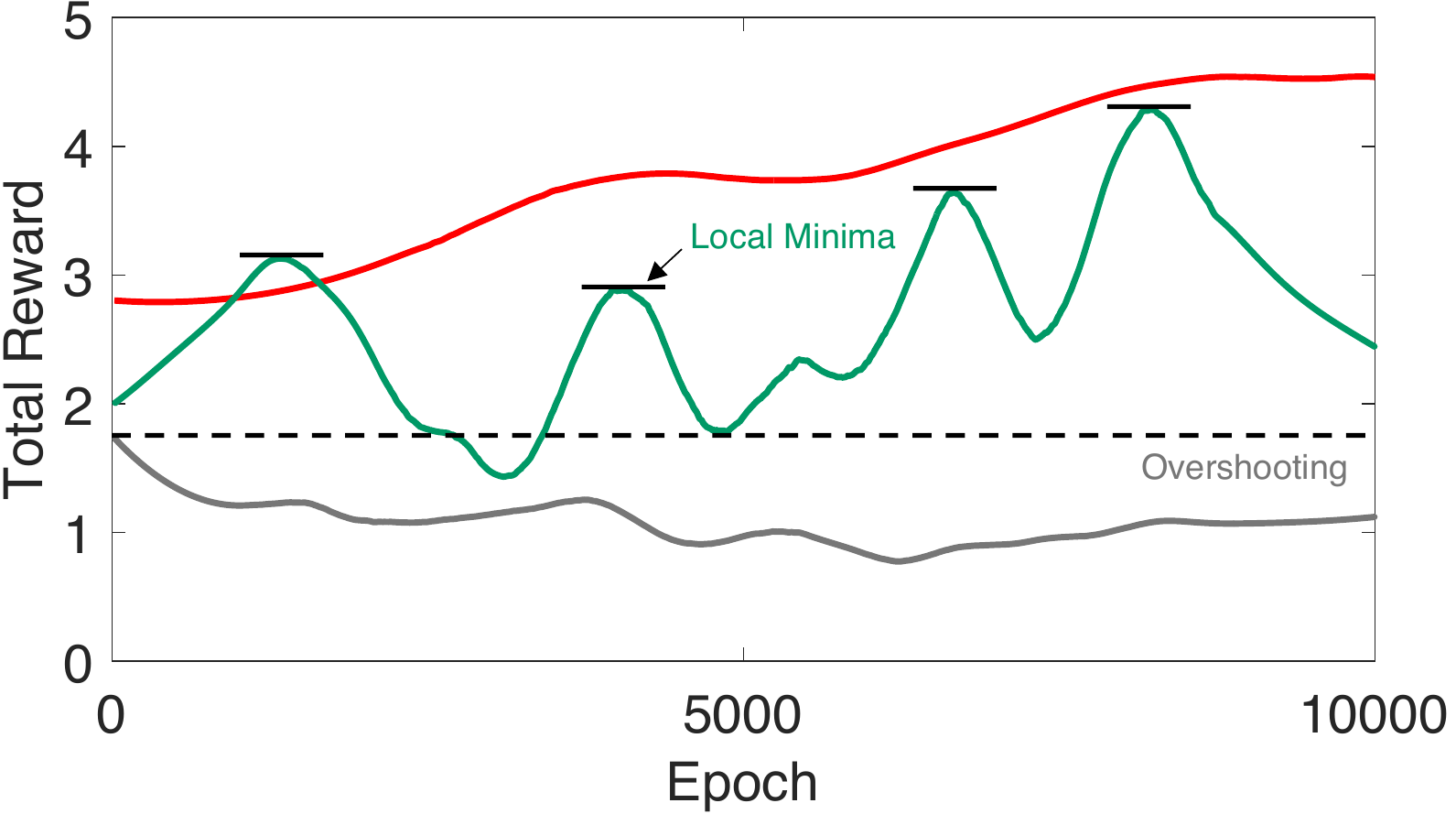}
    \caption{\rev{The outcomes of the agents' policy training according to the chosen learning rate ($0.001$, $0.01$, $0.0001$).}}
    \label{fig:lr}
\end{figure}

\rev{
\subsection{Study on the Impact of Learning Parameters}
In order to empirically evaluate the performance variances due to hyper-parameter setting, our proposed algorithm is evaluated with various learning rate values where the learning rate is one of the major performance-related hyper-parameters~\cite{lv2020deep,ray2019quick}. In RL, model training with overly high learning rates can trigger overshooting, inhibiting the attainment of convergence when policy gradients are utilized to find optimal weights. Conversely, training with low learning rates can introduce excessive numbers of loss function minimization iterations. Moreover, it can engender the risks of local minima realization. Fig.~\ref{fig:lr} illustrates the empirical evaluation results of multi-UAV learning processes when the learning rates are set to $0.01$, $0.001$, and $0.0001$, respectively. It confirms that the reward stably converges to the highest value when the learning rate is $0.001$. It implies that the weights of the rotation gates inside PQC are optimally trained. On the other hand, it is observed that overshooting occurs when the learning rate is $0.01$. Lastly, there are high variances during training when the learning rate is $0.0001$. Furthermore, multiple local minima situations occur.
}

\subsection{Comparison with Classical Neural Networks}

This subsection will determine the hidden layer's dimension of the `CMARL', and investigate quantum performance compared to existing neural networks. Fig.~\ref{fig:with_classical_NN} illustrates the tendency of the reward convergence with the classical neural networks varying from 1 to 128, the ideal model, and the random walk in an ideal environment. Placing an adequate number of neurons in the hidden layers reduces training time with high accuracy, but overfitting problems arise when there are unnecessary increases in neurons or layers. It means that more neural networks with more neurons do not necessarily have better training performance. Among all training benchmark algorithms, our QMACN outperforms in terms of the reward convergence value and speed. Except for QMACN, only classical neural networks with 64 and 128 hidden dimensions obtained higher rewards than the random walk at the end of the training epochs. However, the classical neural network with 128 hidden dimensions has a more unstable training performance than random walk, and fewer rewards than Ideal. Therefore, it is reasonable to set the CMARL's hidden layer level to 64. 
In Fig.~\ref{fig:reward}, the reward convergence trends of all training benchmarks are illustrated over the whole training epochs. {{CMARL}} and {{the random walk}}, which are not based on QMACN, have the inferior performance of training policy in terms of the total reward value than QMACN-based training benchmarks. 
In summary, our proposed QMACN training algorithm enables UAVs to learn the more near-optimal policy than CMARL in a noise-free (\textit{i.e.}, `Ideal') environment.

\subsection{Energy Management for Reliable Mobile Cellular Access}
\begin{figure}[!t]
    \centering
    \includegraphics[width=0.55\linewidth]{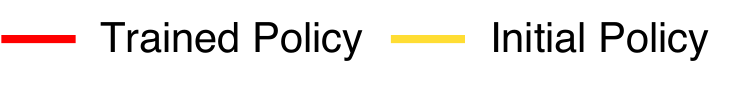}\\
    \subfigure[Total Energy.]{
    \includegraphics[width=0.29\linewidth]{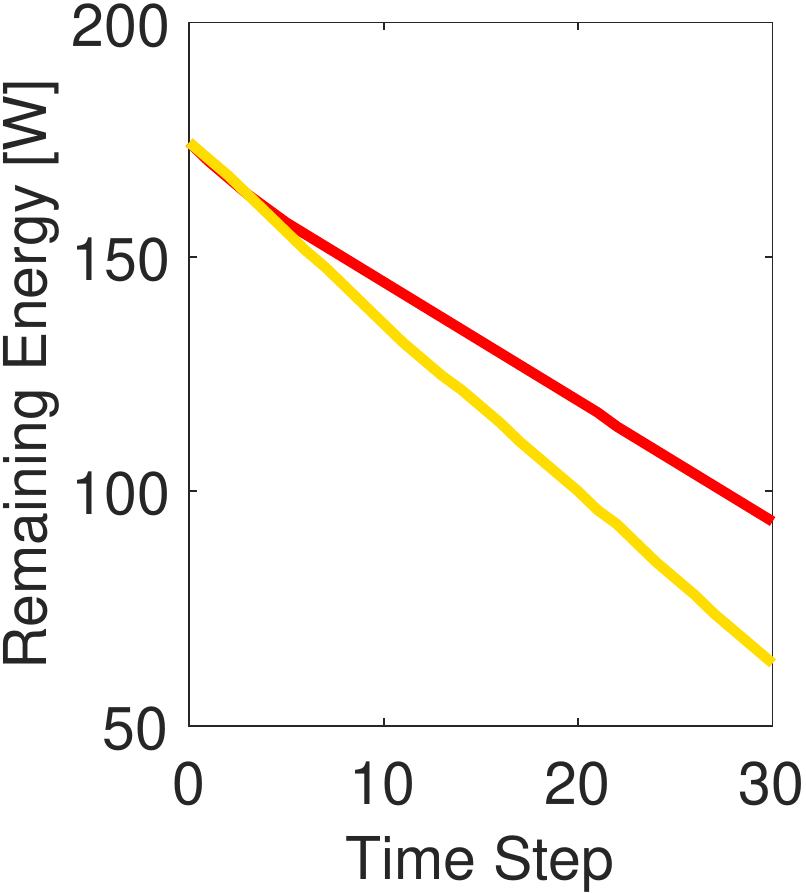}
    \label{fig:total_energy}
    }
    \subfigure[Avg. Energy.]{
    \includegraphics[width=0.29\linewidth]{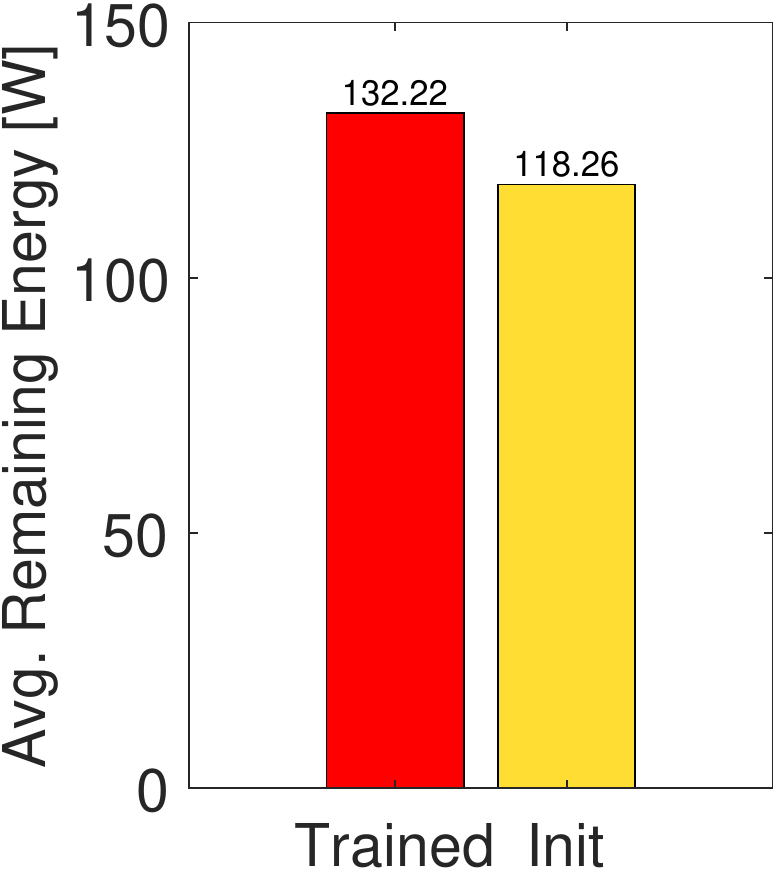}
    \label{fig:average_energy}
    }
    \subfigure[Final Energy.]{
    \includegraphics[width=0.29\linewidth]{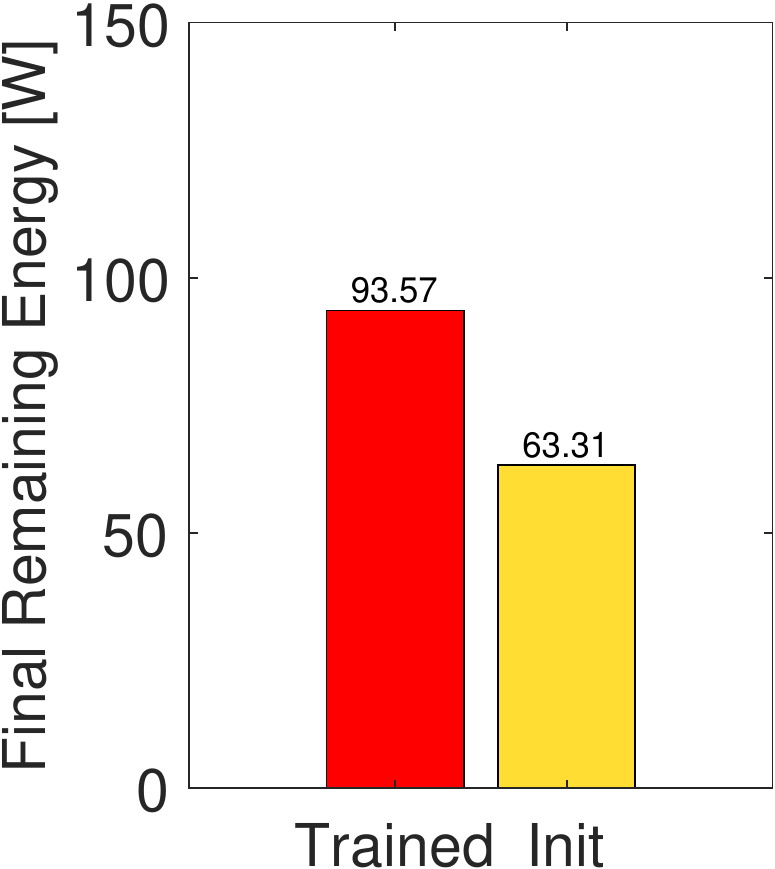}
    \label{fig:final_energy}
    }
    \caption{Energy consumption of all UAV agents before/after training.}
    \label{fig:energy}
\end{figure}

Energy management is vital in UAV networks since UAVs are energy-constrained devices. Therefore, this section investigates the change in the UAV energy consumption after QMACN-based policy training. Fig.~\ref{fig:energy} shows the battery status of all UAVs over the progress of a given time step. It is observed that UAVs with policies trained via QMACN consume less amount of energy than UAVs with untrained policies.
Moreover, UAVs utilizing trained policies have $11.8\,\%$ more energy on average over the entire time step as shown in Fig.~\ref{fig:average_energy} and $47.8\,\%$ more energy than UAVs with untrained policies at the end of the episode as shown in Fig.~\ref{fig:final_energy}.
To recap, the reward function shaped in Sec.~\ref{sec:5-A} successfully trains the policies of UAVs in the direction of maximizing service quality and enhancing network reliability.
However, UAV failures due to a lack of energy cause service inefficiency in mobile cellular access. To deal with this problem, previous work in~\cite{tii202210yun} has demonstrated that MARL-based UAVs adaptively react against the full discharge of a UAV battery by moving into an area covered by a malfunctioning UAV.

\begin{figure}[t!]
    \centering
    \includegraphics[width=0.85\linewidth]{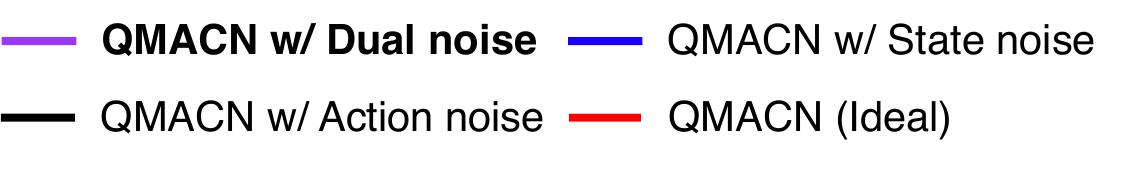}\\
    \subfigure[Support Rate.]{
    \includegraphics[width=0.43\linewidth]{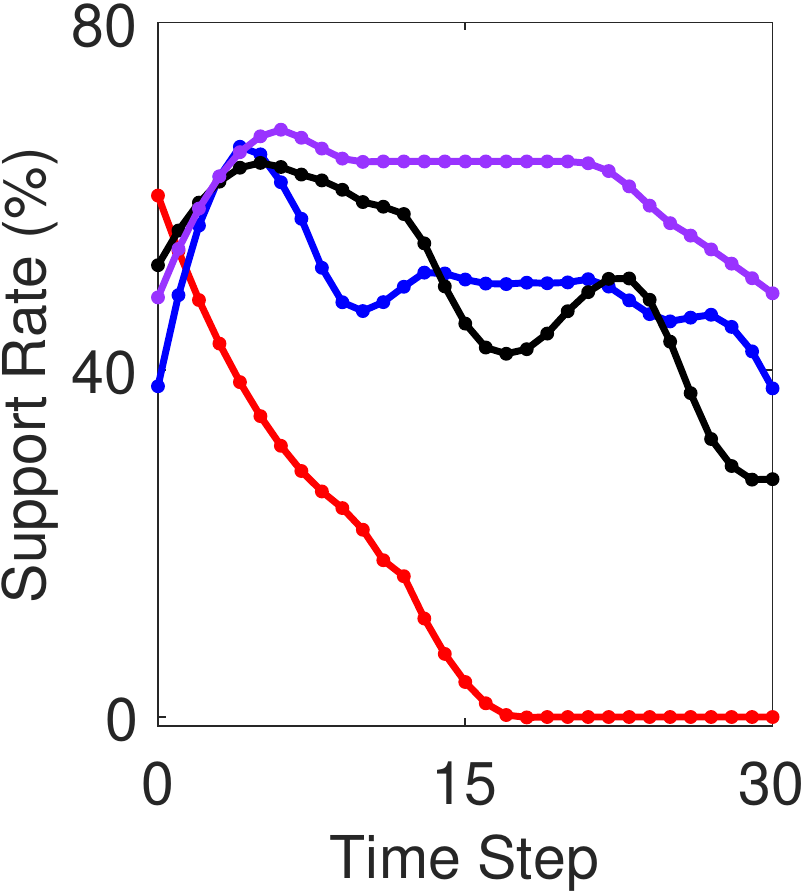}
    \label{fig:inference_support_rate}
    }
    \subfigure[Total QoS.]{
    \includegraphics[width=0.43\linewidth]{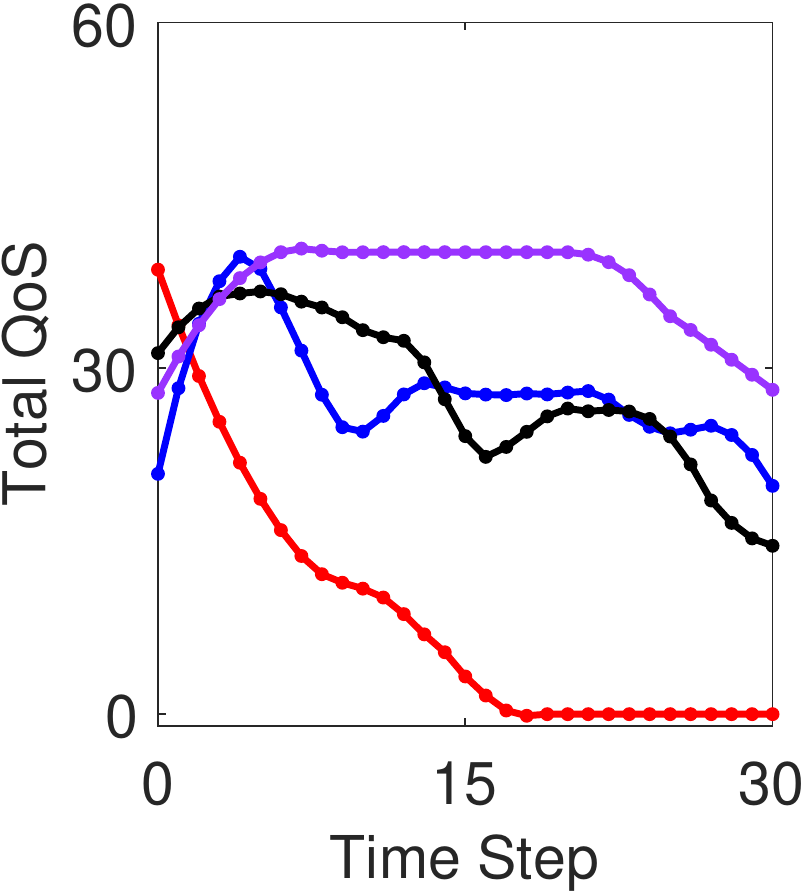}
    \label{fig:inference_qos}
    }
    \\
    \subfigure[Average Support Rate.]{
    \includegraphics[width=0.415\linewidth]{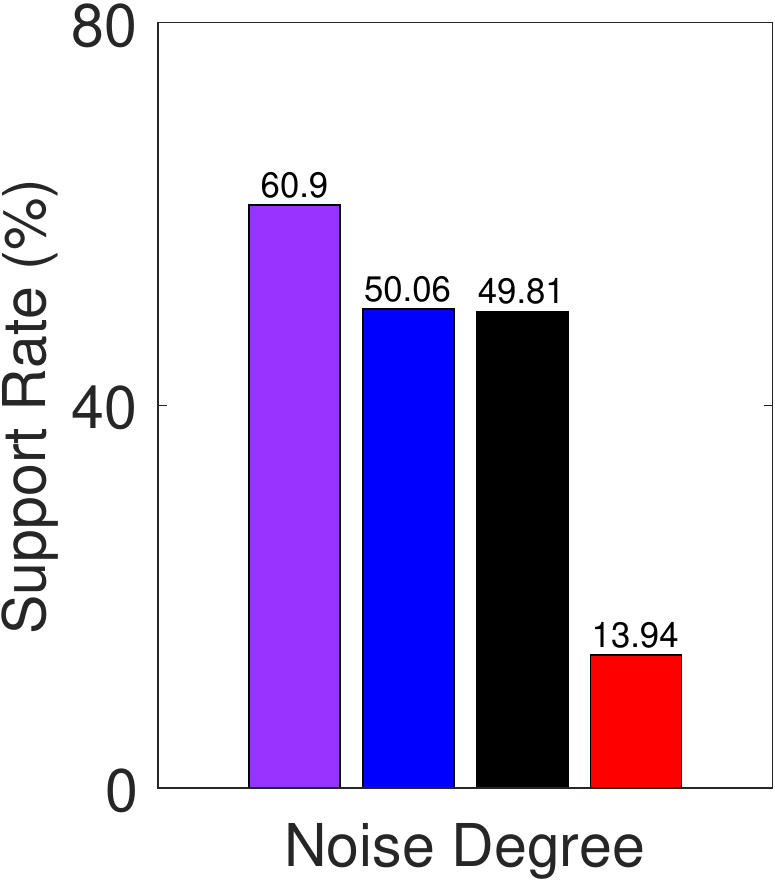}
    \label{fig:inference_support_rate_avg}
    }
    \subfigure[Average Total QoS.]{
    \includegraphics[width=0.415\linewidth]{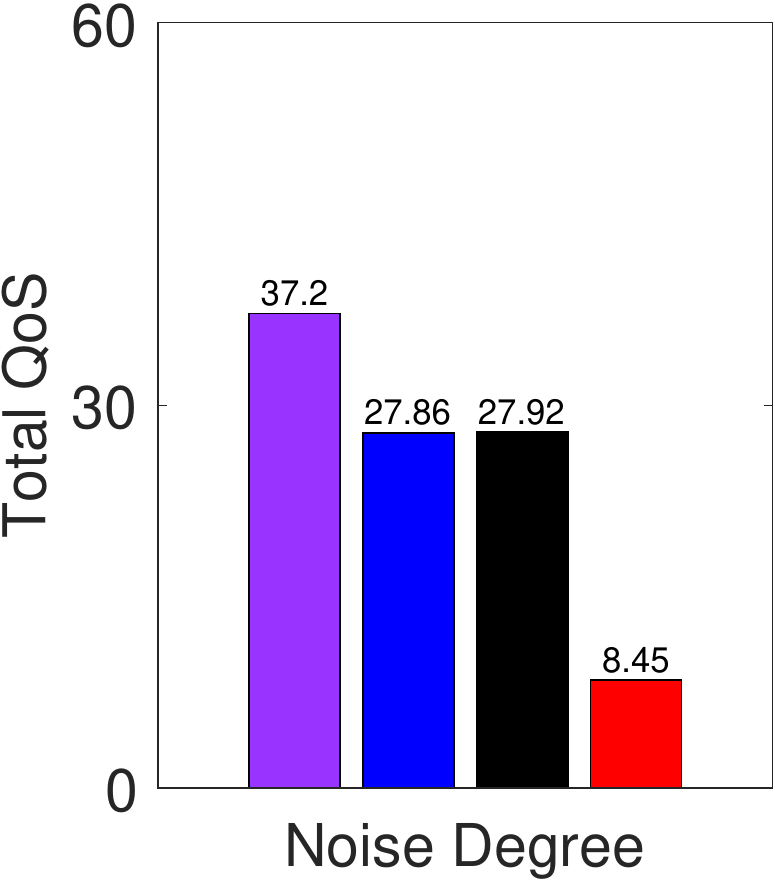}
    \label{fig:inference_qos_avg}
    }
    \caption{Overall support rate and total QoS in the inference process where there are state and action noises.}
    \label{fig:inference}
\end{figure}

\subsection{Impact of Noise on Robustness in Realistic Environment}
After the learning process, all UAVs sequentially make decisions based on the learned policy in the inference process for attesting to the policy's feasibility. We will compare our QMA CNs training in various environments (\textit{i.e.,} ablation study) by inference of 100 iterations. As illustrated in Fig.~\ref{fig:inference}, `w/ Dual noise' has the most highest value regarding support rate (avg.\,60.9\,\%) and QoS (avg.\,37.2) in the entire inference process. `w/ State noise' and `w/ Action noise' constitute the second reliable mobile access with 18\,\% lower support rate and 25\,\% lower QoS than `w/ Dual noise'. 'QMACN' provides the poorest wireless communication service due to environmental noise among all QMACN-based training algorithms. As organized in Table~\ref{tab:inference}, noise-injected QMACNs, including 'QMACN w/ Dual noise', 'QMACN w/ State noise', and 'QMACN w/ Action noise', served more outstanding wireless communication services in a noisy environment than 'QMACN', which trained UAVs' policies without environmental noise. In a nutshell, \revision{the incorporation of increasingly realistic noise considerations leads to more robust policy learning by agents, thereby facilitating the development of effective mobile access systems in real-world scenarios.}

\begin{table}[t!]
\centering         
\caption{Relative performance of noise-injected QMACN w/ Dual noise, w/ State noise, w/ Action noise, over the Ideal QMACN\\in a noisy environment (unit: \%)}
\resizebox{0.88\columnwidth}{!}{\begin{minipage}[h]{0.88\columnwidth}
\centering
\begin{tabularx}{1\linewidth}{c c c}

\toprule[1pt]

\multicolumn{3}{c}{\textbf{\circled{2} : Relative Robustness of QMACN-based UAV Networks}} \\
\midrule[.5pt]
Algorithm & Fig.~\ref{fig:inference_support_rate_avg} & Fig.~\ref{fig:inference_qos_avg}  \\
\cmidrule(lr){1-1} \cmidrule(lr){2-2} \cmidrule(lr){3-3} 

\textbf{w/ Dual noise}
& $46.96\,\%$\; \tikz{
\draw[gray,line width=.3pt] (0,0) -- (1.1,0);
\draw[white, line width=0.01pt] (0,-2pt) -- (0,2pt);
\draw[black,line width=1pt] (0.8642,0) -- (1.0142,0);
\draw[black,line width=1pt] (0.8642,-2pt) -- (0.8642,2pt);
\draw[black,line width=1pt] (1.0142,-2pt) -- (1.0142,2pt);}
& $28.75\,\%$\; \tikz{
\draw[gray,line width=.3pt] (0,0) -- (1.1,0);
\draw[white, line width=0.01pt] (0,-2pt) -- (0,2pt);
\draw[black,line width=1pt] (0.5,0) -- (0.65,0);
\draw[black,line width=1pt] (0.5,-2pt) -- (0.5,2pt);
\draw[black,line width=1pt] (0.65,-2pt) -- (0.65,2pt);} \\

w/ State noise
& $36.12\,\%$\; \tikz{
\draw[gray,line width=.3pt] (0,0) -- (1.1,0);
\draw[white, line width=0.01pt] (0,-2pt) -- (0,2pt);
\draw[black,line width=1pt] (0.6474,0) -- (0.7974,0);
\draw[black,line width=1pt] (0.6474,-2pt) -- (0.6474,2pt);
\draw[black,line width=1pt] (0.7974,-2pt) -- (0.7974,2pt);}
& $19.41\,\%$\; \tikz{
\draw[gray,line width=.3pt] (0,0) -- (1.1,0);
\draw[white, line width=0.01pt] (0,-2pt) -- (0,2pt);
\draw[black,line width=1pt] (0.3132,0) -- (0.4632,0);
\draw[black,line width=1pt] (0.3132,-2pt) -- (0.3132,2pt);
\draw[black,line width=1pt] (0.4632,-2pt) -- (0.4632,2pt);} \\

w/ Action noise
& $35.87\,\%$\; \tikz{
\draw[gray,line width=.3pt] (0,0) -- (1.1,0);
\draw[white, line width=0.01pt] (0,-2pt) -- (0,2pt);
\draw[black,line width=1pt] (0.6424,0) -- (0.7924,0);
\draw[black,line width=1pt] (0.6424,-2pt) -- (0.6424,2pt);
\draw[black,line width=1pt] (0.7924,-2pt) -- (0.7924,2pt);}
& $19.47\,\%$\; \tikz{
\draw[gray,line width=.3pt] (0,0) -- (1.1,0);
\draw[white, line width=0.01pt] (0,-2pt) -- (0,2pt);
\draw[black,line width=1pt] (0.3144,0) -- (0.4644,0);
\draw[black,line width=1pt] (0.3144,-2pt) -- (0.3144,2pt);
\draw[black,line width=1pt] (0.4644,-2pt) -- (0.4644,2pt);} \\

\bottomrule[1pt]
\end{tabularx}
\end{minipage}}
\label{tab:inference}
\end{table}

\subsection{Discussion}
This section will discuss the performance of our proposed QMACN-based training algorithm with the experimental results in Sec.~\ref{sec:5}.

\subsubsection{Quantum Advantage in MARL Environments}
Figs.~\ref{fig:overall_result}--\ref{fig:with_classical_NN} show that our QMACNs based on PQC have comparable policy training performance with fewer parameters when compared to 'CMARL'~\cite{yun2022quantum2}. This empirical advantage lies in memorization, a representative hallmark of PQC~\cite{jerbi2021parametrized}. Preserving knowledge about existing labels when learning new labels helps train policies near-optimally~\cite{yun2022quantum2}. Quantum-based algorithms have sufficient flexibility in this feature, but conventional neural networks which utilize DNN policy do not~\cite{jerbi2021parametrized}. This difference affects the speed or performance of training, as observed in our numerical performance evaluations. In addition, quantum gates employing quantum features (\textit{i.e.,} superposition, entanglement) have the advantage of exponential computational gain over conventional neural networks~\cite{shor1999polynomial}.

\subsubsection{Effects of Noise Injection for Realistic Environment Design}
This paper conducted an ablation study with noise reflection. In general, environmental noise affects the behavior of agents in RL. However, some Gaussian noises improve policy robustness~\cite{he2019parametric} and alleviate the negative effects on the gradient quality by selective noise injection (SNI)~\cite{igl2019generalization}. \revision{This approach not only enables the agent to accomplish desired objectives more effectively but also enhances its resilience to naturally occurring noise.} Our noise models in Eqs.~\eqref{eq:state_noise}--\eqref{eq:action_noise} following Gaussian distributions make UAVs come across abundant experience in environments and prevent overfitting to restricted training environments for a reason in Fig.~\ref{fig:reward}. Even taking some fluctuations into account, benchmarks with noise denoted as `QMACN w/ Dual noise', `QMACN w/ State noise', and `QMACN w/ Action noise' have a faster convergence speed and higher value regarding rewards compared to 'QMACN'. Furthermore, we investigated the model robustness in realistic environments with noise in Fig.~\ref{fig:inference}.

\section{Concluding Remarks}\label{sec:7}
This paper proposes a CTDE-based quantum multi-agent actor-critic networks ({QMACN}) training algorithm for constructing a robust mobile access using multiple UAVs. For the practical use of \revision{a} QC, we adopt a CTDE in overcoming scalability issues \revision{regarding quantum errors during the NISQ era} in order to realize quantum supremacy. Our proposed {QMACN} algorithm verifies the advantage of QMARL with the performance improvements in terms of training speed, wireless service quality, and network reliability in various data-intensive evaluations. Moreover, we validate that the noise injection scheme helps multiple UAVs react to environmental uncertainties, making mobile access more robust. In \revision{summary}, \revision{the} proposed \revision{algorithm named} QMACN \revision{demonstrates} \revision{enhanced} performance in \revision{establishing} cooperative multi-UAV mobile access compared to the conventional training methods with fewer \revision{number of} model parameters, which leads to efficient computing resource management.

\bibliographystyle{IEEEtran}
\bibliography{ref_aimlab, ref_iotj}

\appendices
\revision{
\section{60\,GHz Millimeter-Wave Model}\label{sec:mmwave}

\BfPara{Interference Analysis}
It is essential to consider the impact of interference for performance analysis in wireless networks. In order to conduct the interference analysis for 60\,GHz wireless networks, considering its path loss and antenna pattern is required.
First of all, the 60\,GHz path loss model defined in the IEEE 802.11ad~\cite{maltsev2009path} is as follows:
\begin{equation}
L(d)=32.5 + 20\log_{10}(f_{\mathrm{GHz}})+10\cdot n\cdot \log_{10}(d),
\label{eq:Loss}
\end{equation}
where $f_{\mathrm{GHz}}$, $n$, and $d$ are a carrier frequency in a GHz scale (\textit{i.e.}, $f_{\mathrm{GHz}}=60$), a path loss coefficient ($n=2$), and a distance between transmitter (Tx) and receiver (Rx) in a meter scale. 
Next, the Gaussian 60\,GHz antenna pattern defined in the International Telecommunication Union (ITU) standard document~\cite{sector2019reference} is as follows:
\begin{equation}\label{eq:radiation}
        G_{\mathrm{dBi}}(\varphi, \theta)=
        \begin{cases}
            G_{\mathrm{dBi}}^{\mathrm{max}}-12\Delta^{2}, & 0\leq \Delta < 1\\
            G_{\mathrm{dBi}}^{\mathrm{max}}-12-15\ln{\Delta}, & 1\leq \Delta
        \end{cases}
\end{equation}
where $G_{\mathrm{dBi}}^{\mathrm{max}}$ is a maximum antenna gain, $\varphi$ and $\theta$ are azimuth and elevation angles where $-180^{\circ}\leq \varphi \leq 180^{\circ}$ and  $-90^{\circ}\leq \theta \leq 90^{\circ}$. In addition, $\Delta$ can be obtained with $\Delta \triangleq |\Sigma^{\ddag} \cdot \Sigma^{\dag}|$ 
where 
$\Sigma^{\ddag} \triangleq \cos^{-1}{(\cos{\varphi}\cos{\theta})}$ and 
$\Sigma^{\dag} \triangleq \sqrt{\left[\cos{\left\{\tan^{-1}{\left( \frac{\tan{\theta}}{\sin{\varphi}} \right)}\right\}}/\varphi_3\right]^2 +\left[\sin{\left\{\tan^{-1}{\left( \frac{\tan{\theta}}{\sin{\varphi}} \right)}\right\}}/\theta_3\right]^2}$
where $\varphi_3$ and $\theta_3$ are 3\,dB half-power beamwidth in azimuth and elevation planes; and we assume $\varphi_3 = \theta_3 = 10^{\circ}$~\cite{kim2017feasibility}.
Then, the interference generated by transmissions on wireless link $i$ where $i\neq l,\,i\in \mathcal{I}$ (where $\mathcal{I}$ is a set of wireless links) to the wireless link $l$ can be obtained as follows:
\begin{equation}
    I_{\textrm{mW}}^{i,l}=f_{\textrm{dBm-mW}}\left(\mathcal{G}_{\textrm{dBi}}^{i,\textrm{Tx}}\left(\varphi_1,\varphi_2\right)+\mathcal{P}_{\textrm{dBm}}^{i,\textrm{Tx}}-L\left(d_{\left(i,l\right)}\right)\right),
\label{eq:interference}
\end{equation}
where $\mathcal{P}_{\textrm{dBm}}^{i,\,\textrm{Tx}}$ is the transmit power for the transmission at the link $i$; $L\left(d_{\left(i,l\right)}\right)$ is the path loss, as formulated in Eq.~\eqref{eq:Loss} where $d_{\left(i,l\right)}$ is a distance between the $i$-th Tx and the $l$-th Rx; $\mathcal{G}_{\textrm{dBi}}^{i,\,\textrm{Tx}}$ is an antenna gain between the $i$-th Tx and the $l$-th Rx depending on $\varphi_1$ and $\varphi_2$, which are the angular differences between these Tx and Rx, as presented in Eq.~\eqref{eq:radiation}; $f_{\textrm{dBm-mW}}$ is a function for the translation from dBm to milli-Watt (mW) scales, \textit{i.e.}, $f_{\textrm{dBm-mW}}\triangleq10^{\left(x/10\right)}$.

\begin{table}[t]
\centering
\caption{60\,GHz IEEE 802.11ad MCS Table~\cite{kim2017feasibility}.}
\renewcommand{\arraystretch}{1.3}
\begin{tabular}{c||r|r|r}
\toprule[1pt]   
\textbf{Rx} & \textbf{Available} & \textbf{Maximum} & \textbf{Capacity Estimation} \\
\textbf{Sensitivity} & \textbf{MCS} & \textbf{Data Rates} & \textbf{with Shannon} \\ \midrule
-78\,dBm & MCS0 & 27.5\,Mbps & 1.43\,Gbps \\
-68\,dBm & MCS1 & 385\,Mbps & 2.04\,Gbps \\
-66\,dBm & MCS2 & 770\,Mbps & 2.40\,Gbps \\
-65\,dBm & MCS3 & 962.5\,Mbps & 2.81\,Gbps \\
-64\,dBm & MCS4 & 1155\,Mbps & 3.25\,Gbps \\
-63\,dBm & MCS6 & 1540\,Mbps & 3.74\,Gbps \\
-62\,dBm & MCS7 & 1925\,Mbps & 4.25\,Gbps \\
-61\,dBm & MCS8 & 2310\,Mbps & 5.38\,Gbps \\
-59\,dBm & MCS9 & 2502.5\,Mbps & 7.90\,Gbps \\
-55\,dBm & MCS10 & 3080\,Mbps & 8.57\,Gbps \\
-54\,dBm & MCS11 & 3850\,Mbps & 9.23\,Gbps \\
-53\,dBm & MCS12 & 4620\,Mbps & 43.48\,Gbps \\
\bottomrule[1pt]
\end{tabular}
\label{tab:MCS}
\end{table}

\BfPara{Capacity Calculation With Shannon's Formula}
This paper assumes Shannon's capacity to calculate theoretical data rates in 60\,GHz wireless networks. The Shannon's capacity formula gives the channel capacity $\mathcal{C}_{l}\left(d\right)$ in Eq.~\eqref{eq:Shannon's} at a link $l$ between Tx and Rx with distance $d$, as follows\revision{:}
\begin{equation}\label{eq:Shannon's}
\mathcal{C}_{l}\left(d\right)=\textbf{\textrm{BW}}\cdot\log_{2}\left(\frac{\mathcal{P}_{\textrm{mW}}^{l,\,\textrm{Rx}}\left(d\right)}{n_{\mathrm{mW}}+\sum_{i\in\mathcal{I},\,i\neq l}I_{\textrm{mW}}^{i,l}}+1\right),
\end{equation}
where \textbf{\textrm{BW}} is a 60\,GHz channel bandwidth ($2.16\,\textrm{GHz}$ in IEEE 802.11ad~\cite{kim2017feasibility}); $\mathcal{P}_{\mathrm{mW}}^{l,\mathrm{Rx}}\left(d\right)$ is the received signal stength in a mW scale at the link of the $l$-th Rx; $I_{\textrm{mW}}^{i,l}$ is the interference in Eq.~\eqref{eq:interference}; $n_{\mathrm{mW}}$ is the background noise in a mW scale.
Here, $\mathcal{P}_{\mathrm{dBi}}^{l,\mathrm{Rx}}\left(d\right)$ can be calculated as $\mathcal{P}_{\mathrm{dBi}}^{l,\mathrm{Rx}}\left(d\right)=\underbrace{\mathcal{G}_{\mathrm{dBi}}^{l,\mathrm{Tx}}+\mathcal{P}^{l,\mathrm{Tx}}_{\mathrm{dBm}}}_{\textrm{EIRP}}-L\left(d\right)+\mathcal{G}_{\mathrm{dBi}}^{l,\mathrm{Rx}}$,
    where $\mathcal{G}_{\mathrm{dBi}}^{l,\mathrm{Tx}}$ and $\mathcal{P}^{l,\mathrm{Tx}}_{\mathrm{dBm}}$ are a transmit antenna gain and a transmit power at the Tx of link $l$; $L\left(d\right)$ is a path loss in Eq.~\eqref{eq:Loss}; $\mathcal{G}_{\mathrm{dBi}}^{l,\mathrm{Rx}}$ is a receive antenna gain ($3\,\mathrm{dBi}$ in~\cite{kim2017feasibility}). In~\cite{kim2017feasibility}, the maximum value of equivalent isotropic radiated power (EIRP) in $60$\,GHz channel is set to $43$\,dBm where $\mathcal{G}_{\mathrm{dBi}}^{l,\mathrm{Tx}}=19\,\mathrm{dBi}$ (transmit antenna gain in a dBi scale) and $\mathcal{P}_{\mathrm{dBm}}^{l,Tx}=24\,\mathrm{dBm}$ (transmit power in a dBm scale at link $l$); $n_{\mathrm{mW}}$ is the background noise and $n_{\textrm{mW}}=f_{\textrm{dBm-mW}}\left(k_{\textrm{B}}T_{e}+10\log_{10}\textbf{\textrm{BW}}+\sigma\right)$,
    where $k_{\textrm{B}}T_{e}$ is the noise power spectral density ($-174\textrm{dBm/Hz}$ in~\cite{kim2017feasibility}) and $\sigma$ is an additional loss, \textit{i.e.}, the sum of implementation loss ($10\,\textrm{dB}$ in~\cite{kim2017feasibility}) and noise figure ($5\,\textrm{dB}$ in~\cite{kim2017feasibility}).

\BfPara{Data Rate Computation with IEEE 802.11ad Modulation and Coding Scheme (MCS)}\label{sec:MCS}
Due to transceiver and RF implementation limitations, actual data rates are lower than the rates obtained by Shannon's formula. Therefore, the actual data rate is obtained using the modulation and coding scheme (MCS) in IEEE 802.11ad~\cite{kim2016performance}. Firstly, the Rx signal strength under the consideration of interference is calculated. Then, the actual maximum supportable data rate of the link $l$ is obtained by matching the MCS indices in Table~\ref{tab:MCS}.
}

\newpage

 \begin{IEEEbiographynophoto}{Chanyoung Park} is currently a Ph.D. student in Electrical and Computer Engineering at Korea University, Seoul, Korea, since September 2022. He received the B.S. degree in electrical and computer engineering from Ajou University, Suwon, Korea, in 2022, with an honor (early graduation). 
 \end{IEEEbiographynophoto}

 \begin{IEEEbiographynophoto}{Won Joon Yun} 
 is currently a Ph.D. student in electrical and computer engineering at Korea University, Seoul, Republic of Korea, since March 2021, where he received his B.S. in electrical engineering. 
 \end{IEEEbiographynophoto}

 \begin{IEEEbiographynophoto}{Jae Pyoung Kim} is currently an M.S. student in electrical and computer engineering at Korea University, Seoul, Republic of Korea, since March 2023, where he received his B.S. in electrical engineering. 
 \end{IEEEbiographynophoto}

 \begin{IEEEbiographynophoto}{Tiago Koketsu Rodrigues}
is currently an assistant professor at Tohoku University, Japan. His research interests include artificial intelligence, machine learning, network modeling and simulation, and cloud systems. From 2017 to 2020, he was the Lead System Administrator of \textsc{IEEE Transactions on Vehicular Technology}, overviewing the review process of all submissions and the submission system as a whole. He serves as an Editor for \textsc{IEEE Transactions on Vehicular Technology} and \textsc{IEEE Network}.
 \end{IEEEbiographynophoto}

 \begin{IEEEbiographynophoto}{Soohyun Park} is currently pursuing the Ph.D. degree in electrical and computer engineering at Korea University, Seoul, Republic of Korea. She received the B.S. degree in computer science and engineering from Chung-Ang University, Seoul, Republic of Korea, in 2019. Her research focuses include deep learning algorithms and their applications. 

She was a recipient of the IEEE Vehicular Technology Society (VTS) Seoul Chapter Award in 2019.
 \end{IEEEbiographynophoto}

 \begin{IEEEbiographynophoto}{Soyi Jung} has been an assistant professor at the Department of Electrical of Computer Engineering, Ajou University, Suwon, Republic of Korea, since September 2022. Before joining Ajou University, she was an assistant professor at Hallym University, Chuncheon, Republic of Korea, from 2021 to 2022; a visiting scholar at Donald Bren School of Information and Computer Sciences, University of California, Irvine, CA, USA, from 2021 to 2022; a research professor at Korea University, Seoul, Republic of Korea, in 2021; and a researcher at Korea Testing and Research (KTR) Institute, Gwacheon, Republic of Korea, from 2015 to 2016. She received her B.S., M.S., and Ph.D. degrees in electrical and computer engineering from Ajou University, Suwon, Republic of Korea, in 2013, 2015, and 2021, respectively. 
 \end{IEEEbiographynophoto}

\begin{IEEEbiographynophoto}{Joongheon Kim}
(M'06--SM'18) has been with Korea University, Seoul, Korea, since 2019, where he is currently an associate professor at the School of Electrical Engineering. He received the B.S. and M.S. degrees in computer science and engineering from Korea University, Seoul, Korea, in 2004 and 2006; and the Ph.D. degree in computer science from the University of Southern California (USC), Los Angeles, CA, USA, in 2014. Before joining Korea University, he was a research engineer with LG Electronics (Seoul, Korea, 2006--2009), a systems engineer with Intel Corporation (Santa Clara, CA, USA, 2013--2016), and an assistant professor of computer science and engineering with Chung-Ang University (Seoul, Korea, 2016--2019). 

He serves as an editor for \textsc{IEEE Transactions on Vehicular Technology}, \textsc{IEEE Transactions on Machine Learning in Communications and Networking}, and \textsc{IEEE Communications Standards Magazine}. He is also a distinguished lecturer for \textit{IEEE Communications Society (ComSoc)} and \textit{IEEE Systems Council}. He is an executive director of the Korea Institute of Communication and Information Sciences (KICS). 
 \end{IEEEbiographynophoto}
\end{document}